\documentclass[usenatbib]{mnras}

\usepackage{amsmath}
\usepackage{amssymb}
\usepackage{graphicx}
\usepackage[normalem]{ulem}
\usepackage{graphicx}	% Including figure files
\usepackage{amsmath}	% Advanced maths commands
\usepackage{amssymb}	% Extra maths symbols
\usepackage{rotating}
\usepackage{lscape}
\usepackage{comment}
\usepackage[utf8]{inputenc}
\usepackage[dvipsnames]{xcolor}
\usepackage{siunitx}
\usepackage{verbatim}
\usepackage[english]{babel}
\usepackage[T1]{fontenc}
\usepackage{ae,aecompl}
\usepackage{rotating}
\usepackage{textcomp}
\usepackage{lipsum}
\usepackage{url}
\usepackage[normalem]{ulem}
\usepackage{soul}
\usepackage{lipsum}
\usepackage{hyperref}
\usepackage[switch]{lineno}
\usepackage[dvipsnames]{xcolor}
\title[Optical and $\gamma$-ray emission in BL Lac sources]{The relation between optical and $\gamma$-ray emission in BL Lac sources}

\author[G. La Mura et al.]{G. La Mura\thanks{\textit{E-mail}: glamura@lip.pt},$^1$
 J. Becerra Gonzalez,$^{2,3}$
 G. Chiaro,$^{4}$
 S. Ciroi,$^5$
 J. Otero-Santos$^{2,3}$\\
 \\
 $^1$Laborat\'orio de Instrumenta\c{c}\~ao e F\'{i}sica Experimental de Part\'{i}culas (LIP), Av. Prof. Gama Pinto 2, 1649-003 Lisboa, Portugal\\
 $^2$Instituto de Astrofísica de Canarias (IAC), E-38200 La Laguna, Tenerife, Spain\\
 $^3$Universidad de La Laguna (ULL), Departamento de Astrof\'{i}sica, E-38206 La Laguna, Tenerife, Spain\\
 $^4$Institute of Space Astrophysics and Cosmic Physics, Via A. Corti 12, 20133 Milano, Italy\\
 $^5$Dipartimento di Fisica e Astronomia - Universit\`a di Padova, Via Marzolo 8, 35131 Padova, Italy\\}

\date{Received on July 18, 2022. Accepted on July 19, 2022; in original form on April 11, 2022}
\pubyear{2022}

\begin{document}
\newcommand{\amend}[1]{\textcolor{ForestGreen}{#1}}
\newcommand{\de}{\mathrm{d}}
\maketitle

\begin{abstract}
    The relativistic jets produced by some Active Galactic Nuclei (AGNs) are among the most efficient persistent sources of non-thermal radiation and represent an ideal laboratory for studying high-energy interactions. In particular, when the relativistic jet propagates along the observer's line of sight, the beaming effect produces dominant signatures in the observed spectral energy distribution (SED), from the radio domain up to the highest energies, with the further possibility of resulting in radiation-particle multi-messenger associations. In this work, we investigate the relationships between the emission of $\gamma$ rays and the optical spectra of a sample of AGN, selected from BL Lac sources detected by the {\it Fermi} Large Area Telescope ({\it Fermi}-LAT). We find that there is a close relationship between the optical and gamma-ray spectral indices. Despite all the limitations due to the non-simultaneity of the data, this observation strongly supports a substantial role of Synchrotron-Self Compton (SSC) radiation in a single zone leptonic scenario for most sources. This result simplifies the application of theoretical models to explore the physical parameters of the jets in this type of sources.
\end{abstract}


\section{Introduction}
%We shall definitely refer to LAT \citep[]{LATpaper}.
The presence of intense non-thermal radiation is one of the most general properties of Active Galactic Nuclei (AGN). This spectral component, which is originated in relativistic plasmas threaded by magnetic fields, can be either directly observed in the spectra or leave characteristic signatures in the ionization status of the material distributed at various distances from the source \citep{Urry95}. The most common sites, where the conditions to produce non-thermal radiation are easily met, are the corona of the accretion disk surrounding the central Super Massive Black Hole \citep[SMBH, see e.g.][]{Torricelli05, Liu06} and the relativistic jets that are formed, when the accretion flow is properly coupled with the magnetic fields and the angular momentum of the central SMBH \citep{Blandford77, Algaba17}. Jets, in particular, are very efficient radiation sources because, in spite of being originated within a spatial scale of the order of light-hours, they can propagate through the host galaxies and, in some cases, extend in the inter-galactic medium for hundreds of kiloparsecs \citep{Blandford19}.

Due to the relativistic motion, with bulk Lorentz factors lying approximately in the range $5 \leq \Gamma \leq 50$, the jet radiation is strongly collimated along the direction of motion and, when the beaming cone is oriented towards the observer, its contribution dominates the observed spectrum, making the source appear as a {\it blazar} \citep{Blandford78, Ghisellini93}. Blazars are a special class of radio-loud AGN, characterised by a high degree of variability and polarization and by a spectral energy distribution (SED) that extends from the radio frequencies all the way up to X-ray and $\gamma$-ray energies. While monitoring the $\gamma$-ray sky, the {\it Fermi} Large Area Telescope \citep[{\it Fermi}-LAT,][]{LATpaper} confirmed that blazars are the most common class of persistent extra-galactic $\gamma$-ray sources. Specifically, the third data release of the fourth {\it Fermi}-LAT catalog of $\gamma$-ray sources \citep[4FGL-DR3,][]{Abdollahi22} lists $6659$ entries, out of which $2250$ are classified as blazars, with additional $1493$ blazar candidates of uncertain class (BCU). %report 6659 entries in 4FGL DR3.

Based on their optical spectra, blazars are classically divided into BL Lac type objects, which are dominated by a nearly featureless continuum, with only faint emission lines, sometimes showing host galaxy absorption features, and Flat Spectrum Radio Quasars (FSRQ), which, on the contrary, exhibit prominent emission lines and strong thermal emission from the accretion disk \citep[e.g.,][]{Urry95, Falomo14, Ghisellini11, Sbarrato12}. In both cases the SED displays a characteristic two-hump shape, which is commonly thought to be the result of the combination of a low energy component, spanning from the radio to the UV/X-ray domain and attributed to synchrotron emission, and a high energy one that is generally interpreted as the result of inverse Compton scattering (IC) of low energy photons up to $\gamma$-ray energies by relativistic charged particles. Depending on the frequency where the synchrotron radiation reaches its maximum, blazars are also classified as Low Synchrotron Peaked (LSP, $\nu_{peak}^{syn} \leq 10^{14}\,$Hz), Intermediate Synchrotron Peaked (ISP, $10^{14}\, \mathrm{Hz} < \nu_{peak}^{syn} \leq 10^{15}\,$Hz), High Synchrotron Peaked (HSP, $\nu_{peak}^{syn} \geq 10^{15}\,$Hz) and Extremely High Synchrotron Peaked sources (EHSP, with $\nu_{peak}^{syn} \geq 10^{17}\,$Hz). In general it is found that, while BL Lacs can have SEDs of any class, FSRQ are mostly limited to the LSP and ISP categories. This segregation could be the consequence of a more intense radiation field in FSRQ, with respect to BL Lacs, that powers the high-energy IC component at the expense of the energy of the relativistic particles, thus suppressing high frequency synchrotron emission \citep{Costamante02}.

Since the mechanisms that govern the jet radiation are only broadly understood, a very promising way to investigate jet physics is to collect simultaneous multi-frequency and multi-messenger information that can help constraining our theoretical models, by removing parameter degeneracies. The most useful class of targets for this kind of studies are HSP sources, because the large extension of their SEDs, sometimes reaching up to the TeV domain, offers many spectral windows dominated by non-thermal radiation \citep{Mireia22}. In this work we analysed a sample of $\gamma$-ray blazars with prominent non-thermal emission, trying to assess whether the characteristics of their optical spectra could be used to better identify HSP sources among other source types. Using a technique based on spectral template fitting, we isolated the non-thermal component of the optical spectra and we measured its spectral index. Although we observed statistical trends that actually reflect the SED classification, we did not obtain a strong distinction between the classes. Instead, we observed a close relation between the resulting optical spectral index and the properties of the $\gamma$-ray emission, holding in the majority of the investigated sources, that supports the Synchrotron-Self Compton (SSC) interpretation for this type of objects. The observed properties have relevant implications on the estimate of typical BL Lac jet parameters.

We organize this paper as follows: in \S2 we describe the selection of our sample; in \S3 we discuss the collection and reduction of the optical spectra; in \S4 we present our results and, finally, in \S5 we summarize our conclusions. We use a spectral notation where we assume that power-law spectra are represented by $F_\nu \propto \nu^{-\alpha}$ (or $F_E \propto E^{-\alpha}$). We further adopt a standard $\Lambda$CDM Cosmology, with $H_0 = 69.6\, \mathrm{km\, s^{-1}\, Mpc^{-1}}$, $\Omega_M = 0.286$ and $\Omega_\Lambda = 0.714$ \citep{Bennett14, PlanckColl16}.

\section{Sample selection}
Our investigation aims at the determination of the spectral properties of the non-thermal radiation component. We, therefore, adopted a set of criteria intended to identify sources with strong non-thermal signatures in their optical spectra. We started our work by selecting all objects that are classified as BL Lac type sources from the high Galactic Latitude sample in the second data release of the $4^{th}$ {\it Fermi}-LAT AGN Catalogue \citep[4LAC-DR2,][]{4LACpaper}. We then limited our attention only to the sources with a measured redshift, in order to be able to apply luminosity and energetic arguments and to improve the chances to obtain spectroscopic information. Since the determination of redshift in BL Lac objects is made difficult by the lack of strong spectral features, we took into account only the sources with a redshift value in the NASA Extragalactic Database (NED) service\footnote{\texttt{https://ned.ipac.caltech.edu/}}. However, since some of the reported values are not spectroscopically confirmed and are subject to some degree of uncertainty, in our selection we kept track of the cases where we could obtain a clear spectral confirmation, an uncertain association, or no line detection at all. To obtain the spectra, we cross-correlated the resulting list of sources with different archives of public flux calibrated spectra, including the Sixteenth data release of the Sloan Digital Sky Survey \citep[SDSS-DR16,][]{SDSSDR16}, the second data release of the BAT AGN Spectroscopic Survey \citep[BASS-DR2,][]{BASSpaper}, and the ZBLLac spectroscopic archive \citep{ZBLLACpaper}. For targets without a publicly available spectrum and an optical magnitude $V < 17$, we also performed direct observations with the $1.22\,$m "Galileo" Telescope of the Asiago Astrophysical Observatory, using its B{\&}C spectrograph equipped with a $300\, \mathrm{tr\, mm^{-1}}$ grating. When we found multiple public spectra of the same source, we used the one with the best signal-to-noise ratio (S/N).

\begin{figure*}
    \centering
    \includegraphics[width=0.45\textwidth]{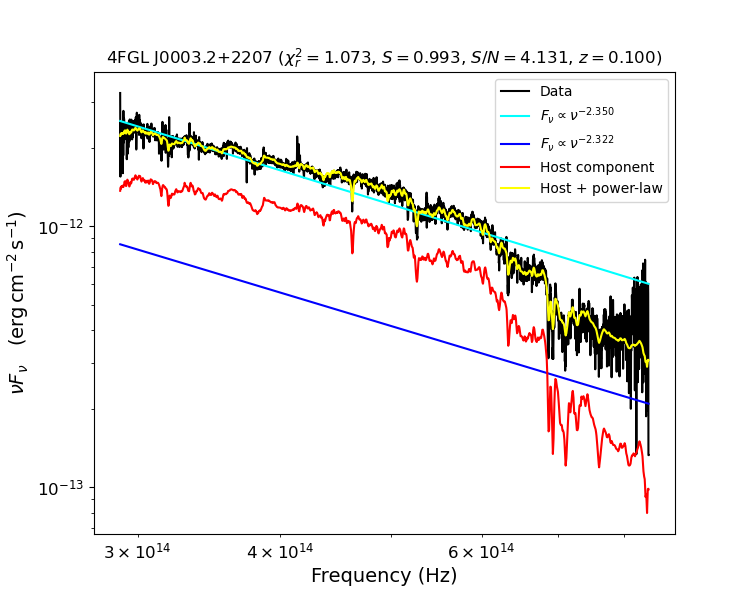}
    \includegraphics[width=0.45\textwidth]{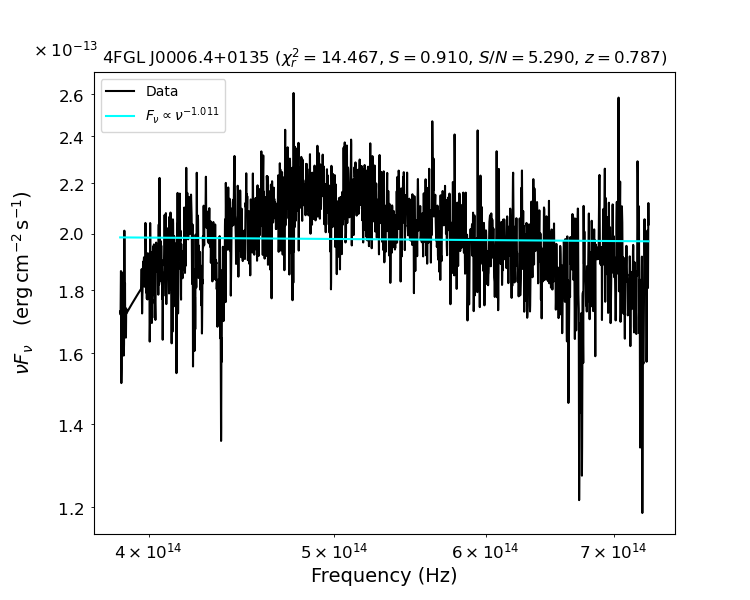}
    \includegraphics[width=0.45\textwidth]{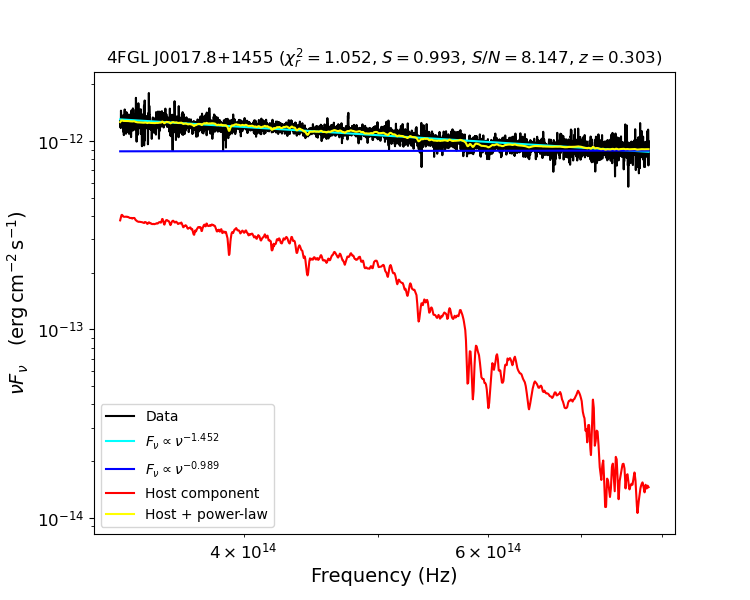}
    \includegraphics[width=0.45\textwidth]{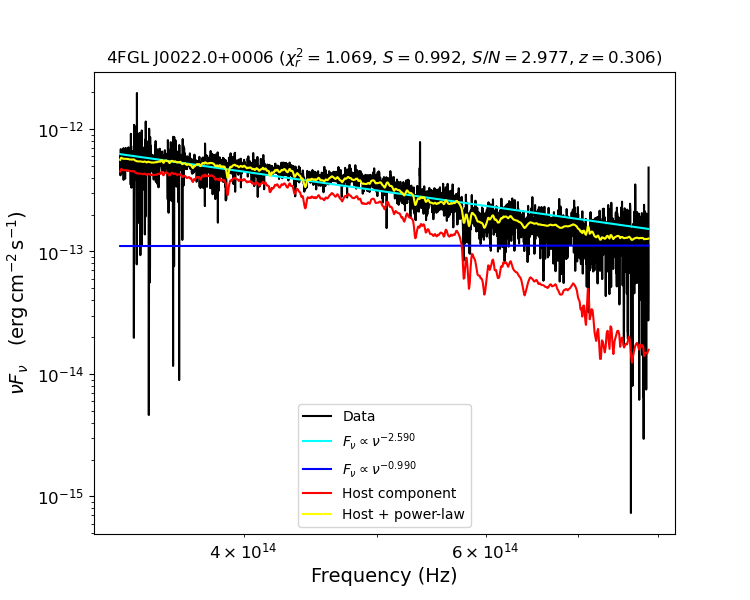}
    \caption{SED representation of the spectra of the first four objects extracted from the sample. The original data are marked with black lines. Cyan lines are used to illustrate a simple power-law fit of the spectrum, while the blue and red lines represent, respectively, the power-law and host galaxy contribution of the multi-component spectral fits, plotted as yellow lines. The illustrated spectra correspond to 4FGL~J0003.2+2207 (accepted), 4FGL~J0003.2+2207 (rejected), 4FGL~J0003.2+2207 (accepted) and 4FGL~J0003.2+2207 (accepted). All the spectra of the sample are available through Google Drive at this \href{https://drive.google.com/file/d/1FVDxmi1JpNGPcyknHDVNSFOl-N97-6tn/view?usp=sharing}{link}. \label{fig01}}
\end{figure*}
As a result, we collected an initial sample of $230$ optical spectra of $\gamma$-ray emitting sources ($5$ observed in Asiago with an average S/N of 7.0, $8$ taken from BASS with average S/N of 4.5, $159$ coming from SDSS with average S/N of 6.3, and $58$ retrieved from ZBLLac with average S/N of 7.3). The main characteristics of this sample are listed in Table~\ref{tab01}. Due to the different origin of our spectroscopic data, we had to perform some preliminary operations, aiming at the construction of a homogeneous data set (see \S3). In addition, although the SEDs of all the selected sources are dominated by non-thermal processes at almost every frequency, in the optical domain we needed to take into account the spectral contributions produced by the host galaxies, which could be present, or even dominant, in a substantial fraction of our sample. Following the example of previous investigations \citep[e.g.,][]{CidFernandes05, Koleva09, Cardoso17}, we addressed this problem through a template based spectral decomposition technique, whose details are described in our data reduction process.

\section{Reduction of optical spectra}
Most of the spectra provided by public databases come in the form of completely flux calibrated data that, in the case of the SDSS and BASS spectra, only needed to be corrected for interstellar reddening. We performed this step applying a standard reddening correction curve \citep{Schlegel98}, normalized according to the color excess maps obtained by \citet{Schlafly11}. This approach is generally sufficient for targets located far away from the Galactic Plane. The spectra obtained from the Asiago Astrophysical Observatory, instead, where reduced in a standard mono-dimensional spectroscopic pipeline that includes bias subtraction, flat field correction, wavelength calibration, through the comparison with He-Fe-Ar arc lamp spectra, taken at each telescope pointing, and flux calibration based on the observation of at least one spectro-photometric standard star per night. The observations were carried out with air masses ranging from 1.1 up to 1.5 and with typical seeing conditions between 2'' and 4''. The last preliminary step was to convert all the spectra into a single \texttt{FITS} file format with $1\,$\AA\ sampling. All these calibration procedures were carried out by means of standard {\it IRAF}~\texttt{v2.16.1} tasks \citep{Tody86, Tody93}.

Since the goal of our analysis is to measure the properties of the non-thermal radiation component and to compare them with the SED classification of our sources, we applied a spectral decomposition technique, which combines a host galaxy template spectrum with a power-law component of type $F_\nu \propto \nu^{-\alpha}$. Following this approach, the SED representation of a spectrum, as a function of frequency $\nu$, becomes:
\begin{equation}
    \nu F_\nu = \nu \left[ k_h \cdot \left( \dfrac{F_\nu^{(h)}}{1\, \mathrm{Jy}} \right) + k_p \cdot \left( \dfrac{\nu}{\nu_0}\right)^{-\alpha} \right], \label{eqnDecom}
\end{equation}
where $k_h$ and $k_p$ are, respectively, the host galaxy template and the power-law component normalization factors,  expressed in Jy units. Using a reference frequency $\nu_0 = 4.996 \cdot 10^{14}\,$Hz, corresponding to $\lambda = 6000\,$\AA, a wavelength covered by the majority of our spectral data set, we looked for the best fit solution through a Levenberg-Marquardt $\chi^2$ minimization algorithm. The host galaxy spectrum was constructed from the template library of \citet{Mannucci01}, by setting the default elliptical galaxy color parameters ($U-B = 0.50$, $B-V = 0.99$, $V-R = 0.59$, $V-I = 1.22$, $V-K = 3.30$, $J-H = 0.69$, $H-K = 0.21$). Each spectrum was then converted into a $\nu \cdot F_\nu$ SED representation and fitted with the host galaxy template artificially moved at the same redshift of the source. The only model free parameters were the host galaxy normalization $k_h$ (fixed to zero, for those cases where its contribution could not be confidently identified), together with the power-law normalization $k_p$ and spectral index $\alpha$. The quality of the fits was estimated by means of residuals assessment, introducing a success rate parameter $S$, defined as the ratio of data points that are within $3\sigma$\ from the model with respect to the total. The spectral fits applied to our sample are illustrated in Fig.~\ref{fig01}.

\begin{figure}
    \centering
    \includegraphics[width=0.44\textwidth]{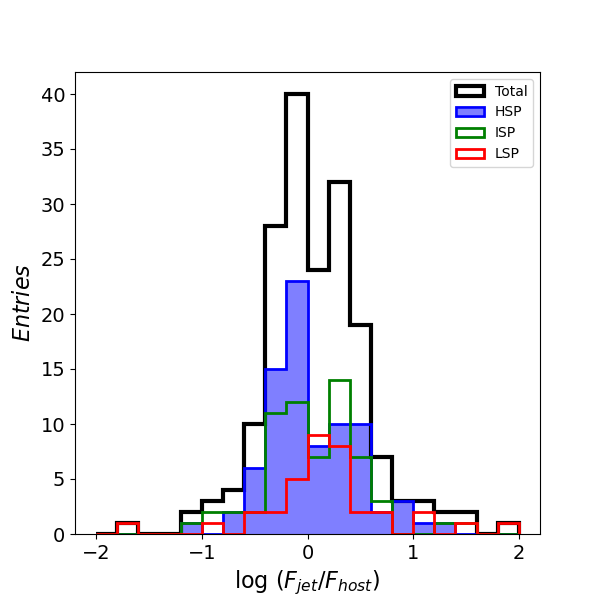}
    \caption{Histograms of the distribution of the logarithm of the flux ratio between jet radiation and host galaxy contribution integrated in the rest frame wavelength range $3500\,${\AA}$ \leqslant \lambda \leqslant 8000\,$\AA\ for spectra with a detectable host component. The thick black histogram is the whole sample, while HSP, ISP and LSP objects are shown, respectively, with blue shaded, green and red histograms. \label{figRatHist}}
\end{figure}
We subsequently inspected the quality of all the fit solutions, marking the spectra whose redshift could be confirmed by the presence of line features, either with good or uncertain association, and rejecting those cases where the fit was not able to reproduce the continuum or where the redshift uncertainty led to severe misinterpretation of the host galaxy component. If the fitting process detected a contribution from the host, we measured its relevance by taking the ratio of the flux attributed to the jet and that attributed to the host, both at the rest frame wavelength $\lambda = 5500\,$\AA\ \citep[similarly to the approach followed by][]{Goldoni21}, as well as by integrating the power-law and the host template components between the rest frame wavelengths of $3500\,$\AA\ and $8000\,$\AA. Typically, we observed acceptable models down to a success rate of $S \approx 0.8$. Unfortunately, the archived spectra of BL Lac and PKS 2155-304 did not come in a form that could be interpreted as the result of combination of a single power-law and one host galaxy component. For these sources, however, we could obtain good quality models using median spectra obtained from the blazar monitoring campaign of the Steward Observatory\footnote{\texttt{https://www.as.arizona.edu/}} \citep{Smith09}. In the case of PKS 2155-304, we combined 299 spectra taken between October 6, 2008 and July 7, 2018. For BL Lac, instead, we used 699 spectra obtained between October 4, 2008 and July 13, 2018. All the spectra were taken with the SPOL/CCD imaging spectro-polarimeter of the 2.3m Bok Telescope of Kitt Peak and the 1.54m Kuiper Telescope of Mt. Bigelow. The spectra covered a wavelength range running from $4000\,$\AA \ to $7550\,$\AA, with a resolution of $16 \leqslant R \leqslant 24$, but we trimmed them at $6650\,$\AA, to avoid the effects of telluric absorptions. Therefore, we finally obtained a sample of $186$ optical spectra that were modeled with acceptable spectral decomposition. We instead discarded from our subsequent considerations the $44$ cases for which we obtained unacceptable fits. This latter circumstance occurred mostly in very low S/N data or in spectra affected by remarkable instrumental or calibration issues.

\begin{table}
    \caption{Average ratios of integrated optical jet-to-host power and spectral indices calculated from our sample of spectra and for the different SED classes. The quoted uncertainties are the standard deviations of the distribution, with the minimum allowed flux ratio trimmed at zero for undetected hosts. \label{tabSpecInd}}
    \begin{center}
    \begin{tabular}{cccc}
    \hline
    \hline
    \vspace{-2mm}
        &     &                      &                 \\
    \vspace{1mm}
    Class & Num. of spectra & $\langle F_{jet}^{int} / F_{host}^{int} \rangle$ & $\langle \alpha_{opt} \rangle$ \\
    \hline
    \vspace{-2mm}
        &     &                      &                 \\
    \vspace{1mm}
    All & 186 & $3.9^{+10.6}_{-3.9}$ & $0.96 \pm 0.62$ \\
    \vspace{1mm}
    HSP &  84 & $2.4^{+4.9}_{-2.4}$ & $0.78 \pm 0.65$ \\
    \vspace{1mm}
    ISP &  65 & $3.3^{+6.8}_{-3.3}$ & $0.98 \pm 0.57$ \\
    \vspace{1mm}
    LSP &  37 & $8.0^{+20.0}_{-8.0}$ & $1.35 \pm 0.40$ \\
    \hline
    \end{tabular}
    \end{center}
\end{table}
\section{Results and discussion}
\subsection{Host galaxy contribution}
Applying the aforementioned template based decomposition to the optical spectra collected in our sample, we are practically able to separate the stellar contribution from the non-thermal component and to estimate their properties. Our analysis aims at measuring the relative strengths of the jet and host galaxy spectral components and to determine the spectral index of the non-thermal power-law contribution. For the flux ratios, we took two different measurements: the ratio of specific fluxes at rest frame wavelength $5500\,$\AA\ \citep[using the same approach as][]{Goldoni21}, and an integrated flux ratio, taken between $3500\,$\AA\ and $8000\,$\AA\ rest frame wavelengths. We compare the two measurements in Appendix~\ref{secCompareRatios} and, since the results are fairly similar, we focus our discussion on the integrated flux ratio.  The results of our measurements are summarized in Table~\ref{tabSpecInd}, for the whole sample, and reported in detail in Table~\ref{tab02}. As it is shown in Fig.~\ref{figRatHist}, our sample spans a range of objects that runs from host-dominated up to jet-dominated sources. In the majority of cases, the flux attributed to the host galaxy and the jet component, in the optical domain, remains comparable, with a slight trend for HSP and ISP sources to appear more host dominated than the LSP ones. We can also observe a similar trend of sources being more host dominated in the low redshift part of the sample.

Providing a proper interpretation of this result, on the basis of our data set, is not a straightforward task. Generally speaking, we can expect the formation of BL Lac type optical spectra from two fundamentally different configurations: a very powerful jet, whose boosted emission dominates over all the other AGN components, and a jet launched by a source powered by a radiatively inefficient accretion process, therefore lacking of the characteristic optical emission lines and accretion related continuum common to other types of AGNs. In our case, we need to take into account that, by selecting BL Lac sources with a commonly accepted redshift estimate, we may have systematically suppressed a large portion of objects with strong optical jet contributions. This is well reflected by the fact that the average estimate of optical luminosity seen in our sample is $\langle \nu L_\nu^{opt} \rangle \approx 10^{45}\, \mathrm{erg\, s^{-1}}$, while a bright \textit{quasar} such as 3C~273 exceeds $10^{46}\, \mathrm{erg\, s^{-1}}$ in the optical domain. Whatever is the origin of the BL Lac optical spectrum, an increasing weight of the host in H/ISP sources, with respect to LSP ones, suggests that jets with lower optical radiation power can more easily carry high energy particles, moving the peak of their SED towards higher frequencies and, thus, resulting in an intrinsic \textit{blazar sequence} effect \citep[e.g.][]{Fossati98, Ghisellini98, Ghisellini08, Prandini22}. These considerations, however, would be more properly addressed in a more homogeneous sample that, taking into account the role of atmospheric seeing, of spectrograph aperture and of target angular scale, may provide a better control of the amount of AGN and host galaxy flux collected by the observations, particularly in low redshift targets.

\begin{figure}
    \centering
    \includegraphics[width=0.45\textwidth]{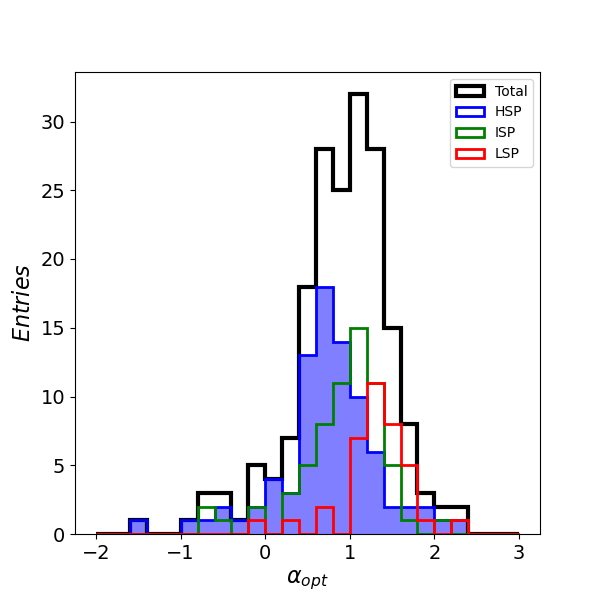}
    \caption{Histograms of the distribution of measured spectral indices for the non-thermal component in the optical spectra fitted with acceptable models. The solid black histogram represents the spectral index distribution of the whole sample, while we also plot separately objects classified as HSP (blue shaded histogram), ISP (green histogram) and LSP (red histogram) BL lac sources. \label{figIndHist}}
\end{figure}
\subsection{Spectral index measurements}
Once the host galaxy and the power-law components of the spectra have been separated, we can estimate the spectral index of the non-thermal contribution. Since, by definition of SED classification, we can identify HSP objects as those whose synchrotron emission peak lies above $\nu_{peak}^{syn} = 10^{15}\,$Hz, which is in the UV domain, we expect that the optical spectra of HSP sources should be characterized by a $F_\nu \propto \nu^{-\alpha_{opt}}$, with $\alpha_{opt} \leq 1$. As it is shown in Fig.~\ref{figIndHist} and in Table~\ref{tab02}, the situation that we find in our sample is somehow different from this simple prediction. Indeed, if we consider the spectral indices measured in objects belonging to different SED classes, there is no sharp separation among them, but rather a smooth transition between object types. This apparent contradiction, however, is easily interpreted in view of some very simple considerations. In first place, we can estimate that the spectral index measurement in the optical domain is affected by an intrinsic uncertainty $\Delta \alpha_{opt} \geqslant 0.2$, which tends to be larger in objects whose optical spectrum is more host-dominated. This uncertainty might be regarded as a systematic effect in our method, descending from the assumptions that all host galaxies can be represented by a single template spectrum and that the host and jet components are the only radiation sources to be accounted for. Secondly, the presence of ISP objects, which can also have a blue or UV synchrotron peak, accounts for a large fraction of the non-HSP sources with $\alpha_{opt} \leq 1$. Finally, we need to take into account the intrinsic variability of sources, which, due to the non-simultaneous nature of the data that lead to a SED classification, can well explain the observation of spectral indices that do not agree with the expected values.

\begin{figure*}
    \centering
    \includegraphics[width=0.95\textwidth]{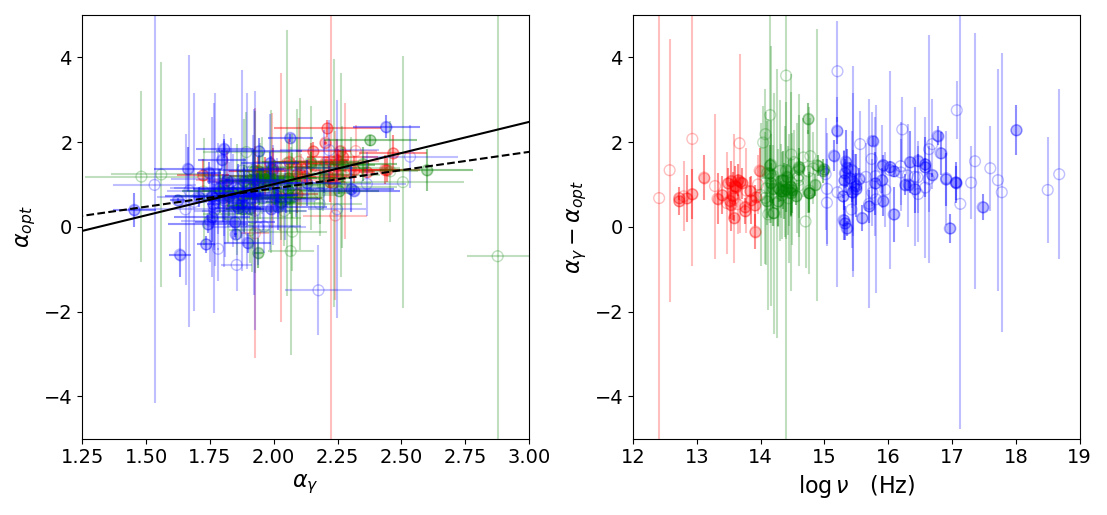}
    \caption{The relation between the optical non-thermal spectral index $\alpha_{opt}$ and the $\gamma$-ray power-law index $\alpha_\gamma$ (left panel) and the difference $\alpha_\gamma - \alpha_{opt}$ plotted as a function of logarithmic frequency (right panel). The symbol colors are used to distinguish among HSP (blue), ISP (green) and LSP (red) objects. We plot as open circles the sources with $\Delta \alpha_{opt} > 0.6$, while filled circles represents the ones with $\Delta \alpha_{opt} \leq 0.6$. The dashed and solid lines correspond, respectively, to the linear fits of Eq.~(\ref{eqnWhole}) and Eq.~(\ref{eqnBest}). \label{figSpecInd}}
\end{figure*}
In spite of the above considerations, we can appreciate a statistical trend of increasing spectral index (i.e. softer average spectra), while moving through the HSP, ISP and LSP classes. This result, summarized in Table~\ref{tabSpecInd}, is consistent with our natural expectations and it supports the reliability of spectral decomposition techniques to extract and analyse the non-thermal emission of these objects. As a consequence, we can investigate the relation existing between the non-thermal component of the optical spectra and the corresponding $\gamma$-ray spectra. It is well known that the $\gamma$-ray photon index correlates with the position of the synchrotron peak in frequency \citep[e.g.][]{4LACpaper}. So far, however, the presence of thermal contributions in the optical domain has made it very difficult to perform direct comparisons of spectral indices between the optical and the $\gamma$-ray energy domains. From our sample, instead, we can infer a substantial degree of linear correlation among the optical spectral index $\alpha_{opt}$ and its $\gamma$-ray counterpart $\alpha_\gamma$. The situation, summarized in Fig.~\ref{figSpecInd}, is well represented by a linear relation described by:
\begin{equation}
    \alpha_{opt} = (0.862 \pm 0.110) \cdot \alpha_\gamma - (0.818 \pm 0.445), \label{eqnWhole}
\end{equation}
with weighted correlation coefficient $R = 0.743$ and null-hypothesis probability $p_0 = 2.71 \cdot 10^{-10}$ (i.e. statistical significance at $6.3\, \sigma$\ level), for the whole sample, and by:
\begin{equation}
    \alpha_{opt} = (1.469 \pm 0.224) \cdot \alpha_\gamma -(1.931 \pm 0.919), \label{eqnBest}
\end{equation}
with weighted correlation coefficient $R = 0.751$ and null-hypothesis probability $p_0 = 1.54 \cdot 10^{-8}$ (i.e. $5.7\, \sigma$), for the subset of $98$ objects that were fitted with $\Delta \alpha_{opt} \leq 0.6$ (corresponding to $3$ times the systematic uncertainty estimated for the method). The observed correlations are particularly suggestive, if compared to the situation that is obtained by simply fitting a power-law model to the optical data, without accounting for the host component (more details in Appendix~\ref{secFitNoHost}). Another interesting result, also presented in Fig.~\ref{figSpecInd}, is that the above correlation results in a situation where the power-law fits of the $\gamma$-ray spectra appear to be softer than the associated optical non-thermal emission by a value that lies mostly close to $1.0$. This circumstance points towards a tight relationship between the population of particles that explain the non-thermal component of the optical spectra and the ones which are responsible for the $\gamma$-ray emission. The existence of a similar relationship between the observed spectral indices is a characteristic prediction of SSC based models. However, the systematic difference between the optical and $\gamma$-ray energy regions requires further investigation to be properly interpreted.

\subsection{Constraints from the observed SED properties}
It is generally well established that the broad band SED of blazars can be described in terms of a combination of synchrotron radiation and inverse-Compton scattered photons. In the case of FSRQ, the intense radiation fields produced by efficiently radiating accretion processes are thought to be responsible of a strong Compton cooling effect on the most energetic radiating particles, which results in a lower energy of the synchrotron photons and in a relatively more prominent Compton component. The environment expected in BL Lac type objects, on the contrary, lacks such additional contributions and the photons that are scattered are most likely originated by the jet itself. In particular, when the particles that are responsible for the synchrotron emission are the same on which the IC process produces the high energy radiation, the source operates in SSC mode. It can be demonstrated that this scenario predicts a strong relation between the spectral indices measured in the optical domain and in the $\gamma$-ray energies, as seen in our data. For this reason, we set up a modeling framework to test the agreement of SSC predictions with the properties observed by {\it Fermi}-LAT and in our optical spectra. A full description of the model implementation is given in Appendix~\ref{secModel}.

\begin{figure}
    \centering
    \includegraphics[width=0.45\textwidth]{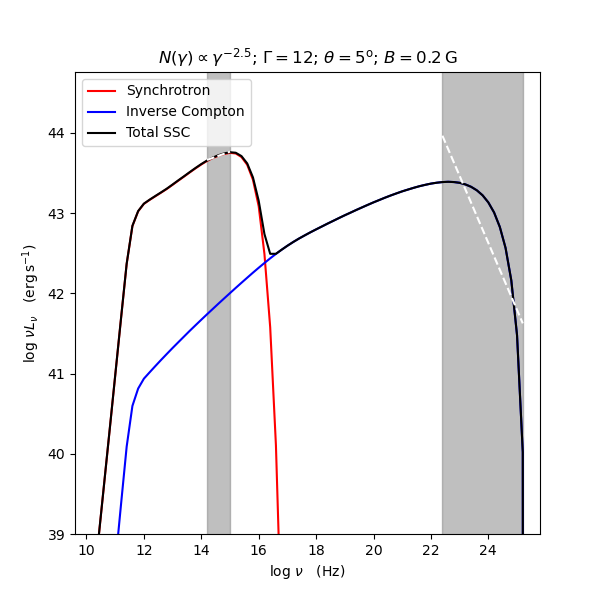}
    \caption{Example SSC model SED for a blob of relativistic plasma with total particle density $N = 3 \cdot 10^5\, \mathrm{cm^{-3}}$, particle energy distribution $N(\gamma) \propto \gamma^{-2.5}$ ($4 \leq \gamma \leq 2.5 \cdot 10^4$), average magnetic field $B = 0.2\,$G, bulk Lorentz factor $\Gamma = 12$ and inclination over the line of sight $\vartheta = 5^\mathrm{o}$. The Doppler factor is $\delta = 11.462$. We represent the synchrotron emission as a red line, the IC component as a blue line and the resulting spectrum as the black continuous line. The grey shaded areas represent the frequency ranges used to apply spectral power-law fits, shown as white dashed lines. \label{figSSCmodel}}
\end{figure}

The mathematical expressions used to evaluate the spectrum can be addressed by means of numerical integration codes. In our case, we combined the spectra predicted by Eq.~(\ref{eqnSynSpec}) and Eq.~(\ref{eqnICspec}) in a \textit{python} program that calculates the expected SSC emission, given a set of input parameters that describe the average magnetic field, the size of the emitting region, the particle density and their energy distribution. The predicted spectrum is then Doppler shifted and boosted to the observer frame, by taking into account the effects of relativistic beaming for a given inclination $\vartheta$ and bulk Lorentz factor $\Gamma$. An example of this type of model is presented in Fig.~\ref{figSSCmodel}.

The most evident signature of single-zone leptonic SSC models is a tight relationship between the spectral indices measured in the synchrotron and the IC components. This relation descends from the fact that the energy distribution of a single particle species controls both the synchrotron emission and the frequency redistribution via Compton scattering. The simplest possible relation occurs when the scattering operates in the Thomson regime. In this case, the two spectral indices are expected to be approximately equal. For relativistic particles, on the contrary, the high energy part of the spectrum tends to be curved and softened, due to the onset of scattering in the Klein-Nishina regime, where the cross-section drops very quickly below the Thomson value. As a consequence, the power-law approximation of the $\gamma$-ray spectrum appears to be softer than the corresponding fit on the optical synchrotron component. Assuming that the $\gamma$-ray power-law index measured by {\it Fermi}-LAT corresponds to an energy window of $0.1\, \mathrm{GeV} \leqslant E \leqslant 100\, \mathrm{GeV}$, this effect can be clearly appreciated in Fig.~\ref{figSSCmodel}, where the spectral indices of the best fit power-laws in the optical and in the $\gamma$-ray domain are, respectively, $\alpha_{opt} = 0.87$ and $\alpha_\gamma = 1.83$, leading to $\alpha_\gamma - \alpha_{opt} = 0.96$. In principle, another effect that can account for a softer $\gamma$-ray index can be the suppression of the high energy spectrum due to Extragalactic Background Light opacity effects \citep[EBL,][]{Dominguez11,Franceschini19}. However, this effect is expected to have only a secondary role in our sample, because the opacity to photons with $E = 100\,$GeV is estimated as $\tau_E = 0.865$ for $z = 1.0$ \citep{Saldana21}. The spectra that gave rise to acceptable fits are distributed in the redshift range $0.03 \leqslant z \leqslant 1.83$, with a median redshift of $0.276$. In particular, only $8$ objects of the sample, with $z > 1$, resulted in acceptable fits. Since the majority of our sample lies at much smaller redshift and the photons statistics is typically dominated by lower energies, the fact that our $\gamma$-ray emitting BL Lac sources present a tight relationship between the the optical and $\gamma$-ray spectral indices, with an average difference of $\langle \alpha_\gamma - \alpha_{opt} \rangle = 1.14 \pm 0.62$, strongly supports the idea that a leptonic SSC scenario, with a Klein-Nishina cut-off, can be an appropriate interpretation of the underlying radiation mechanism.

\begin{figure*}
\begin{center}
    \includegraphics[width=0.95\textwidth]{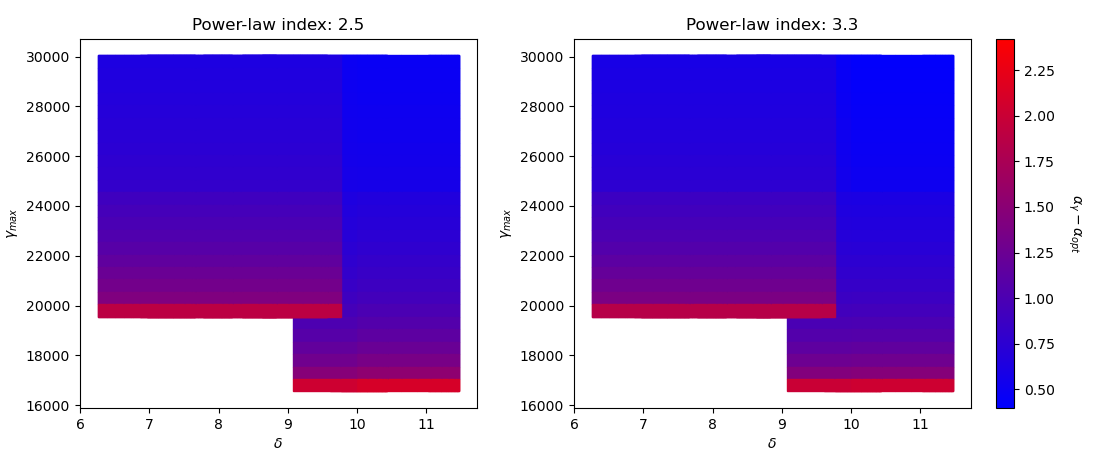}
\end{center}
\caption{The $\gamma$-ray-to-optical spectral index difference predicted by SSC models as a function of the jet Doppler factor and of the maximum Lorentz factor of the emitting particles, for power-law energy distributions of type $N_e(\gamma) \propto \gamma^{-2.5}$ (left panel) and $N_e(\gamma) \propto \gamma^{-3.3}$ (right panel). \label{figGridModels}}
\end{figure*}
In the case of SSC emission, several parameters describing the jet physics can be constrained through the application of relatively simple models. We know, indeed, that leptonic radiation models operate in relatively low magnetic fields ($B \leqslant 1\,$G), in order to be consistent with the X-ray spectral properties. The fast observed variability, after accounting for relativistic corrections, implies emitting regions with sizes in the order of $R \approx 10^{15}\,$cm. Therefore, the measured luminosity and the ratio between the synchrotron component, with respect to the Comptonized radiation, can be used to estimate the density of the radiating particles. As it can be deduced from Eq.~(\ref{eqnJsyn}) and Eq.~(\ref{eqnJic}), the expected synchrotron and IC emissivities are respectively proportional to $N$ and to $N^2$, resulting in emitting particle densities in the range $10^5\, \mathrm{cm^{-3}} \leqslant N \leqslant 10^7\, \mathrm{cm^{-3}}$, in order to sustain optical luminosities between $10^{43}\, \mathrm{erg\, s^{-1}}$ and $10^{46}\, \mathrm{erg\, s^{-1}}$. Interestingly, these densities are typical of the AGN environment at the parsec scale and consistent with the jet parameter range suggested by other studies \citep[e.g.][]{Mireia22}. Using a magnetic field value of $B = 0.2\,$G in the SSC framework of Appendix~\ref{secModel}, we computed a grid of $17940$ models, for jet inclinations varying from $5^\mathrm{o}$ to $7.5^\mathrm{o}$ in steps of $0.5^\mathrm{o}$, bulk Lorentz factors in the range $5 \leqslant \Gamma \leqslant 15$ (in steps of $0.5$) and particle energy distributions with power-law indices running from $2.5$ to $3.5$ in steps of $0.2$. We further adopted a minimum particle Lorentz factor of $4$ and we let the maximum Lorentz factor decrease from an input threshold of $2.98 \cdot 10^4$ in steps of $500$, until the modeled particle distributions were no longer able to sustain $\gamma$-ray emission. Considering that the spectral windows, used to define the optical and $\gamma$-ray domain in our observer frame, correspond to a rest-frame frequency range that varies up to a factor of $\sim 3$, due to the redshift distribution of our sample, we verified through our models that the observed low-to-high energy index difference can be fairly well reproduced by the chosen parameter set. Investigating the predictions of different models, we can see that the expected difference is slightly affected by the Doppler factor and, more strongly, by the maximum Lorentz factor attained by the emitting particles. The use of various energy indices for the particle energy distribution has only a secondary effect, as illustrated in Fig.~\ref{figGridModels}.

All these estimates represent a suggestive insight in the physics of AGN jets, but, at this stage, they only give a set of reference values that apply, on average, at least to the case of $\gamma$-ray emitting BL Lac objects. The use of SSC models provides a natural interpretation to most of the observed spectral index differences, but it can not explain the few cases where the optical index appears softer than the average $\gamma$-ray power-law index. This situation, however, is only observed in a handful of cases ($6$ sources out of $186$) and it appears to be of real concern only for 4FGL J1548.3+1456 (NVSS J154824+145702). These cases can be seen in objects with very soft optical spectra, which might have been affected by peculiar short term variability effects at the time of their observation. More detailed modeling of specific sources will certainly require carefully selected simultaneous data, that can likely turn the range of parameter estimates seen in our sample into more precise physical measurements.

\section{Conclusions}
The properties of AGN jets are an important field of research in modern astrophysics. Thanks to their characteristically large energies and collimated emission, they act both as very effective high energy laboratories and as cosmic beacons that trace galaxy evolution activity across the Universe. However, their range in power, variability and spectral properties, together with remarkable uncertainties in their structure and acceleration mechanisms, require extensive multi-frequency monitoring campaigns, in order to constrain their theoretical interpretation and to remove the implied parameter degeneracies. The best targets for this type of studies are objects with jet dominated SEDs, which are most commonly found among HSP BL Lac type sources.

In this study, we selected a sample of $\gamma$-ray emitting sources, where we expected the jet contribution to be dominant, and we analysed the relation existing between the radiation emitted at optical and $\gamma$-ray frequencies. We followed a template spectral fitting technique that uses a simple elliptical host galaxy spectrum and a power-law component to reproduce the data. This technique was sufficient to model the spectra of $186$ sources out of $230$, consistent with the expectation that BL Lac sources are hosted in rather typical elliptical galaxies \citep{Urry00}. The relative strength of the jet with respect to the host, at rest frame optical wavelengths, tends to be in the range $0.1 \leqslant f_{ratio} \leqslant 10$, with few more extreme cases. Typically, the host component appears to be stronger in HSP/ISP objects, with respect to LSP ones, and more prominent in low redshift sources than in the high redshift regime, though there are notable exceptions. However, the related uncertainties, which are only estimated by taking into account the normalization errors (i.e. neglecting the uncertainty on power-law indices), are too large to draw strong conclusions on this point.

By removing the contribution of the host galaxy stellar population, we first investigated the spectral index of the resulting non-thermal radiation, interpreted as non-thermal jet emission, as a function of the source SED classification. Although we found a well defined trend of spectral hardening, going from the LSP, through the ISP and HSP SED classes, we could not observe a clear separation among them, mainly due to the intermediate properties of the ISP class and to the possibility that non simultaneous SED reconstruction and optical spectroscopic observations could lead to mismatched optical properties, in a population of variable targets. Instead, we observed a strong correlation between the spectral indices inferred from power-law fits to the optical and the $\gamma$-ray data collected by {\it Fermi}-LAT. The existence of this correlation is predicted in the framework of SSC radiation models and its observation in our sample supports the idea that leptonic SSC is a likely interpretation for the average $\gamma$-ray emission of BL Lac type sources.

The spectral index distribution observed in our sample exhibits a striking relation between the optical and the $\gamma$-ray band, which cannot be pointed out, if the jet radiation is not isolated from the host galaxy stellar contribution. After we properly take into account the host galaxy component, the remaining non-thermal spectrum appears to be almost ubiquitously harder than the $\gamma$-ray spectrum. The spectral index difference is averagely close to $1$, with only a small number of sources showing remarkable differences. If interpreted in the framework of SSC emission, this circumstance suggests that $\gamma$-ray emitting BL Lac type AGNs span over a common range of tightly related jet parameters. The properties observed in our sample, in the assumption of typical structural values for this class of sources, are consistent with the gas densities characteristic of the close environment of AGNs and are well explained by particle energy distributions that extend up to Lorentz factors $\gamma_e^{max} \sim 10^4$.

The application of SSC models to the optical and $\gamma$-ray spectral properties of BL Lac type sources can well explain the average properties observed in the majority  of our sources. On the contrary, these models cannot easily explain the cases when the optical spectral index turns out to be softer than the $\gamma$-ray one. This situation is, however, uncommon and not particularly significant in our sample. Strong variability effects or the possible contribution of additional emission components, that are not seen in the average properties of the radiation mechanism, are very likely playing an important role in these cases. Our results, therefore, suggest that SSC emission is suitable to justify the average properties of $\gamma$-ray emitting BL Lac sources and it can be used to perform detailed analyses of jet parameters, if applied to properly executed simultaneous multi-frequency campaigns.

\section*{Data policy statement}
Most of the data used in this study were extracted from the credited data sources. The $\gamma$-ray properties are the ones listed in 4LAC-DR2, while the optical spectra were selected from the BASS, SDSS and ZBLLAC public archives. The new spectra collected at the Asiago Astronomical Observatory can be obtained with a request to the corresponding author or downloaded from Google Drive at the following \href{https://drive.google.com/file/d/1Fkvun-YbbvMrImZQDpiV1w4vKrkC8GML/view?usp=sharing}{link}.

%\clearpage
%%%TABLE 1: SAMPLE DESCRIPTION
\begin{table*}
\caption{List of sources included in the sample. The table columns report, respectively: the source name in 4FGL catalogue, a common counterpart name, the equatorial coordinates of the counterpart (J2000.0), the adopted source redshift $z$, the SED class, the success rate of the spectral fit $S$, the origin of the spectral data (ASG for the Asiago Observatory, BAS for the BASS survey, SDS for the SDSS DR16 spectra ZBL for the ZBLLac repository, and STO for spectra that were obtained from the Steward Observatory), a flag reporting whether the spectrum has identified lines (yes / no / uncertain) and one informing whether the spectral fit was deemed acceptable. \label{tab01}}
\begin{tabular}{llcccccccc}
\hline
\hline
Object name        & Counterpart                & R.A.     & Dec.       & $z$   & SED & $S$   & Source & Line Id. & Accept \\
\hline
4FGL J0003.2+2207  & 2MASX J00032450+2204559    & 00:03:25 & +22:04:56  & 0.100 & LSP & 0.993 & SDS         & Yes      & Yes    \\
4FGL J0006.4+0135  & NVSS J000626+013611        & 00:06:27 & +01:36:10  & 0.787 & ISP & 0.910 & ZBL         & No       & No     \\
4FGL J0017.8+1455  & GB6 J0017+1450             & 00:17:37 & +14:51:02  & 0.303 & ISP & 0.993 & SDS         & Yes      & Yes    \\
4FGL J0022.0+0006  & RX J0022.0+0006            & 00:22:01 & +00:06:58  & 0.306 & HSP & 0.992 & SDS         & Yes      & Yes    \\
4FGL J0028.8--0112 & PKS 0026-014               & 00:29:01 & --01:13:42 & 0.083 & ISP & 0.986 & SDS         & Yes      & Yes    \\
4FGL J0035.2+1514  & RX J0035.2+1515            & 00:35:15 & +15:15:04  & 1.073 & ISP & 0.988 & SDS         & No       & Yes    \\
4FGL J0040.4--2340 & PMN J0040-2340             & 00:40:25 & --23:40:01 & 0.213 & ISP & 0.597 & ZBL         & Yes      & Yes    \\
4FGL J0050.7--0929 & PKS 0048-09                & 00:50:41 & --09:29:05 & 0.635 & ISP & 0.902 & ZBL         & No       & No     \\
4FGL J0056.3--0935 & TXS 0053-098               & 00:56:20 & --09:36:30 & 0.103 & ISP & 0.987 & SDS         & Yes      & Yes    \\
4FGL J0059.3--0152 & RX J0059.3-0150            & 00:59:17 & --01:50:18 & 0.144 & HSP & 0.990 & SDS         & Yes      & Yes    \\
4FGL J0101.0--0059 & NVSS J010058-005547        & 01:00:58 & --00:55:48 & 1.240 & ISP & 0.976 & SDS         & No       & Yes    \\
4FGL J0109.1+1815  & MG1 J010908+1816           & 01:09:08 & +18:16:08  & 1.038 & ISP & 0.987 & SDS         & No       & No     \\
4FGL J0111.4+0534  & 1RXS J011130.5+053612      & 01:11:30 & +05:36:27  & 0.347 & ISP & 0.988 & SDS         & Yes      & Yes    \\
4FGL J0115.8+2519  & RX J0115.7+2519            & 01:15:46 & +25:19:53  & 0.376 & HSP & 0.985 & SDS         & Yes      & Yes    \\
4FGL J0121.8--3916 & NVSS J012152-391547        & 01:21:53 & --39:15:44 & 0.390 & HSP & 0.866 & ZBL         & Yes      & Yes    \\
4FGL J0152.6+0147  & PMN J0152+0146             & 01:52:40 & +01:47:17  & 0.080 & HSP & 0.997 & ASG         & Yes      & Yes    \\
4FGL J0201.1+0036  & MS 0158.5+0019             & 02:01:06 & +00:34:00  & 0.299 & HSP & 0.984 & SDS         & Yes      & Yes    \\
4FGL J0203.7+3042  & NVSS J020344+304238        & 02:03:44 & +30:42:38  & 0.760 & LSP & 0.975 & SDS         & Yes      & No     \\
4FGL J0220.8--0841 & RX J0220.8-0842            & 02:20:48 & --08:42:50 & 0.525 & HSP & 0.982 & SDS         & Unc.     & Yes    \\
4FGL J0237.6--3602 & RBS 0334                   & 02:37:34 & --36:03:28 & 0.411 & HSP & 0.923 & ZBL         & Unc.     & Yes    \\
4FGL J0238.7+2555  & NVSS J023853+255407        & 02:38:54 & +25:54:07  & 0.584 & HSP & 0.922 & ZBL         & Yes      & Yes    \\
4FGL J0242.9+0045  & FIRST J024302.9+004627     & 02:43:03 & +00:46:27  & 0.409 & ISP & 0.993 & SDS         & Unc.     & Yes    \\
4FGL J0250.6+1712  & RGB J0250+172              & 02:50:38 & +17:12:09  & 0.243 & HSP & 0.708 & ZBL         & Yes      & Yes    \\
4FGL J0301.0--1652 & PMN J0301-1652             & 03:01:17 & --16:52:45 & 0.278 & LSP & 0.962 & ZBL         & Yes      & Yes    \\
4FGL J0303.4--2407 & PKS 0301-243               & 03:03:27 & --24:07:11 & 0.266 & HSP & 0.769 & ZBL         & No       & No     \\
4FGL J0304.5--0054 & RX J0304.5-0054            & 03:04:34 & --00:54:05 & 0.511 & HSP & 0.975 & SDS         & Unc.     & Yes    \\
4FGL J0305.1--1608 & PKS 0302-16                & 03:05:15 & --16:08:17 & 0.312 & HSP & 0.451 & ZBL         & Yes      & Yes    \\
4FGL J0316.2--2608 & RBS 0405                   & 03:16:15 & --26:07:57 & 0.443 & HSP & 0.830 & ZBL         & Unc.     & Yes    \\
4FGL J0325.5--5635 & 1RXS J032521.8-563543      & 03:25:24 & --56:35:45 & 0.060 & HSP & 0.656 & BAS         & Yes      & No     \\
4FGL J0326.2+0225  & 1H 0323+022                & 03:26:14 & +02:25:15  & 0.147 & HSP & 0.975 & ZBL         & Yes      & Yes    \\
4FGL J0340.5--2118 & PKS 0338-214               & 03:40:36 & --21:19:31 & 0.223 & LSP & 0.822 & ZBL         & No       & Yes    \\
4FGL J0353.0--6831 & PKS 0352-686               & 03:52:58 & --68:31:17 & 0.087 & HSP & 0.887 & BAS         & Yes      & Yes    \\
4FGL J0428.6--3756 & PKS 0426-380               & 04:28:40 & --37:56:20 & 1.105 & LSP & 0.740 & ZBL         & Unc.     & No     \\
4FGL J0505.6+0415  & MG1 J050533+0415           & 05:05:35 & +04:15:55  & 0.423 & HSP & 0.484 & ZBL         & Yes      & No     \\
4FGL J0543.9--5531 & 1RXS J054357.3-553206      & 05:43:57 & --55:32:07 & 0.273 & HSP & 0.874 & ZBL         & No       & Yes    \\
4FGL J0550.5--3216 & PKS 0548-322               & 05:50:41 & --32:16:16 & 0.680 & HSP & 0.773 & ZBL         & No       & No     \\
4FGL J0558.0--3837 & EXO 0556.4-3838            & 05:58:06 & --38:38:32 & 0.302 & HSP & 0.682 & ZBL         & Unc.     & Yes    \\
4FGL J0710.4+5908  & 1H 0658+595                & 07:10:30 & +59:08:20  & 0.125 & HSP & 0.983 & ASG         & Yes      & Yes    \\
4FGL J0727.1+3734  & SDSS J072659.51+373423.0   & 07:27:00 & +37:34:23  & 0.791 & ISP & 0.985 & ZBL         & Unc.     & Yes    \\
4FGL J0731.9+2805  & RGB J0731+280              & 07:31:53 & +28:04:33  & 0.248 & ISP & 0.985 & SDS         & Yes      & Yes    \\
4FGL J0733.7+4110  & GB6 J0733+4111             & 07:33:47 & +41:11:20  & 0.195 & ISP & 0.991 & SDS         & No       & Yes    \\
\hline
\end{tabular}
\end{table*}
\section*{Acknowledgements}
We gratefully acknowledge the discussion and suggestions of the referee, which led to substantial improvement of this work. This work was funded by Funda\c{c}\~ao para a Ci\^encia e a Tecnologia, under project PTDC/FIS-PAR/4300/2020.

The results presented in this work are based on data published by the \textit{Fermi}-LAT Collaboration. The data are made available thanks to the support of several agencies and institutions. These include the National Aeronautics and Space Administration and the Department of Energy in the United States, the Commissariat \`a l’Energie Atomique and the Centre National de la Recherche Scientifique / Institut National de Physique Nucl\'eaire et de Physique des Particules in France, the Agenzia Spaziale Italiana and the Istituto Nazionale di Fisica Nucleare in Italy, the Ministry of Education, Culture, Sports, Science and Technology (MEXT), High Energy Accelerator Research Organization (KEK) and Japan Aerospace Exploration Agency (JAXA) in Japan, and the K. A. Wallenberg Foundation, the Swedish Research Council and the Swedish National Space Board in Sweden. Additional support for science analysis during the operations phase is gratefully acknowledged from the Istituto Nazionale di Astrofisica in Italy and the Centre National d’\'Etudes Spatiales in France.

Data from the Steward Observatory spectropolarimetric monitoring project were used. This program is supported by Fermi Guest Investigator grants NNX08AW56G, NNX09AU10G, NNX12AO93G, and NNX15AU81G.

Funding for the Sloan Digital Sky Survey IV has been provided by the Alfred P. Sloan Foundation, the U.S. Department of Energy Office of Science, and the Participating Institutions. 

SDSS-IV acknowledges support and resources from the Center for High Performance Computing  at the University of Utah. The SDSS website is www.sdss.org.

SDSS-IV is managed by the Astrophysical Research Consortium for the Participating Institutions of the SDSS Collaboration including the Brazilian Participation Group, the Carnegie Institution for Science, Carnegie Mellon University, Center for Astrophysics | Harvard \& Smithsonian, the Chilean Participation Group, the French Participation Group, Instituto de Astrof\'isica de Canarias, The Johns Hopkins University, Kavli Institute for the Physics and Mathematics of the Universe (IPMU) / University of Tokyo, the Korean Participation Group, Lawrence Berkeley National Laboratory, Leibniz Institut f\"ur Astrophysik Potsdam (AIP),  Max-Planck-Institut f\"ur Astronomie (MPIA Heidelberg), Max-Planck-Institut f\"ur Astrophysik (MPA Garching), Max-Planck-Institut f\"ur Extraterrestrische Physik (MPE), National Astronomical Observatories of China, New Mexico State University, New York University, University of Notre Dame, Observat\'ario Nacional / MCTI, The Ohio State University, Pennsylvania State University, Shanghai Astronomical Observatory, United Kingdom Participation Group, Universidad Nacional Aut\'onoma de M\'exico, University of Arizona, University of Colorado Boulder, University of Oxford, University of Portsmouth, University of Utah, University of Virginia, University of Washington, University of Wisconsin, Vanderbilt University, and Yale University.

\bibliographystyle{mnras}
\bibliography{biblio}
%\clearpage
%%%TABLE 1: SAMPLE DESCRIPTION
\begin{table*}
\contcaption{}
\begin{tabular}{llcccccccc}
\hline
\hline
Object name        & Counterpart                & R.A.     & Dec.       & $z$   & SED & $S$   & Source & Line Id. & Accept \\
\hline
4FGL J0740.9+3203  & LEDA 1979979               & 07:41:06 & +32:05:44  & 0.179 & HSP & 0.995 & SDS         & Yes      & Yes    \\
4FGL J0749.2+2314  & RX J0749.2+2313            & 07:49:14 & +23:13:17  & 0.175 & HSP & 0.992 & SDS         & Yes      & Yes    \\
4FGL J0754.7+4823  & GB1 0751+485               & 07:54:46 & +48:23:51  & 0.377 & LSP & 0.773 & SDS         & No       & No     \\
4FGL J0757.1+0956  & PKS 0754+100               & 07:57:07 & +09:56:35  & 0.266 & LSP & 0.214 & ZBL         & No       & No     \\
4FGL J0758.9+2703  & SDSS J075846.99+270515.5   & 07:58:47 & +27:05:16  & 0.099 & LSP & 0.988 & SDS         & Yes      & Yes    \\
4FGL J0809.6+3455  & B2 0806+35                 & 08:09:39 & +34:55:37  & 0.082 & HSP & 0.988 & SDS         & Yes      & Yes    \\
4FGL J0811.4+0146  & OJ 014                     & 08:11:27 & +01:46:52  & 1.148 & LSP & 0.831 & ZBL         & Unc.     & No     \\
4FGL J0812.9+5555  & NVSS J081251+555422        & 08:12:51 & +55:54:22  & 0.383 & HSP & 0.982 & SDS         & Unc.     & Yes    \\
4FGL J0818.2+4222  & S4 0814+42                 & 08:18:16 & +42:22:45  & 0.331 & LSP & 0.828 & SDS         & No       & No     \\
4FGL J0818.4+2816  & GB6 J0818+2813             & 08:18:27 & +28:14:03  & 0.225 & ISP & 0.989 & SDS         & Yes      & Yes    \\
4FGL J0819.4+4035  & GB6 J0819+4037             & 08:19:26 & +40:37:44  & 0.389 & LSP & 0.997 & SDS         & Unc.     & Yes    \\
4FGL J0827.0--0708 & PMN J0827-0708             & 08:27:06 & --07:08:46 & 0.247 & ISP & 0.645 & ZBL         & Yes      & No     \\
4FGL J0831.8+0429  & PKS 0829+046               & 08:31:49 & +04:29:39  & 0.174 & LSP & 0.864 & SDS         & Unc.     & No     \\
4FGL J0837.3+1458  & RGB J0837+149              & 08:37:25 & +14:58:20  & 0.278 & HSP & 0.939 & SDS         & Yes      & Yes    \\
4FGL J0842.5+0251  & NVSS J084225+025251        & 08:42:26 & +02:52:53  & 0.425 & ISP & 0.992 & SDS         & Unc.     & Yes    \\
4FGL J0847.2+1134  & RX J0847.1+1133            & 08:47:13 & +11:33:50  & 0.198 & ISP & 0.990 & SDS         & Yes      & Yes    \\
4FGL J0850.5+3455  & RX J0850.5+3455            & 08:50:36 & +34:55:23  & 0.145 & HSP & 0.994 & SDS         & Yes      & Yes    \\
4FGL J0854.0+2753  & SDSS J085410.16+275421.7   & 08:54:10 & +27:54:22  & 0.494 & ISP & 0.991 & SDS         & Unc.     & Yes    \\
4FGL J0857.7+0137  & RX J0857.8+0135            & 08:57:50 & +01:35:30  & 0.281 & HSP & 0.992 & SDS         & Yes      & Yes    \\
4FGL J0901.4+4542  & NVSS J090208+454433        & 09:02:08 & +45:44:33  & 0.288 & ISP & 0.993 & SDS         & Yes      & Yes    \\
4FGL J0910.8+3859  & FBQS J091052.0+390202      & 09:10:52 & +39:02:02  & 0.199 & LSP & 0.992 & SDS         & Yes      & Yes    \\
4FGL J0911.7+3349  & MG2 J091151+3349           & 09:11:48 & +33:49:17  & 0.456 & HSP & 0.992 & SDS         & Unc.     & Yes    \\
4FGL J0916.7+5238  & RX J0916.8+5238            & 09:16:52 & +52:38:28  & 0.190 & HSP & 0.990 & SDS         & Yes      & Yes    \\
4FGL J0917.3--0342 & NVSS J091714-034315        & 09:17:15 & --03:43:14 & 0.308 & HSP & 0.977 & ZBL         & Yes      & Yes    \\
4FGL J0930.5+4951  & 1ES 0927+500               & 09:30:38 & +49:50:26  & 0.187 & HSP & 0.995 & SDS         & Yes      & Yes    \\
4FGL J0937.9--1434 & NVSS J093754-143350        & 09:37:55 & --14:33:50 & 0.287 & ISP & 0.720 & ZBL         & Yes      & Yes    \\
4FGL J0940.4+6148  & RX J0940.3+6148            & 09:40:22 & +61:48:26  & 0.211 & HSP & 0.994 & SDS         & Yes      & Yes    \\
4FGL J0942.3+2842  & NVSS J094223+284413        & 09:42:23 & +28:44:14  & 0.366 & HSP & 0.993 & SDS         & Yes      & Yes    \\
4FGL J0945.7+5759  & GB6 J0945+5757             & 09:45:42 & +57:57:48  & 0.229 & LSP & 0.993 & SDS         & Yes      & Yes    \\
4FGL J0952.8+0712  & SDSS J095249.57+071329.9   & 09:52:50 & +07:13:30  & 0.574 & ISP & 0.877 & ZBL         & Yes      & Yes    \\
4FGL J0955.1+3551  & 1RXS J095508.2+355054      & 09:55:08 & +35:51:01  & 0.834 & ISP & 0.991 & SDS         & No       & Yes    \\
4FGL J0959.4+2120  & RX J0959.4+2123            & 09:59:30 & +21:23:21  & 0.365 & ISP & 0.993 & SDS         & Yes      & Yes    \\
4FGL J1003.6+2605  & PKS 1000+26                & 10:03:42 & +26:05:13  & 0.929 & LSP & 0.992 & SDS         & No       & Yes    \\
4FGL J1008.8--3139 & PKS 1006-313               & 10:08:51 & --31:39:05 & 0.534 & LSP & 0.632 & ZBL         & Yes      & No     \\
4FGL J1012.3+0629  & NRAO 350                   & 10:12:13 & +06:30:57  & 0.727 & LSP & 0.992 & SDS         & Unc.     & Yes    \\
4FGL J1021.9+5123  & MS 1019.0+5139             & 10:22:13 & +51:24:00  & 0.142 & ISP & 0.992 & SDS         & Yes      & Yes    \\
4FGL J1023.8+3002  & RX J1023.6+3001            & 10:23:40 & +30:00:58  & 0.433 & HSP & 0.992 & SDS         & Unc.     & Yes    \\
4FGL J1024.8+2332  & MG2 J102456+2332           & 10:24:54 & +23:32:34  & 0.165 & LSP & 0.987 & SDS         & Yes      & Yes    \\
4FGL J1028.3+3108  & TXS 1025+313               & 10:28:18 & +31:07:34  & 0.240 & ISP & 0.987 & SDS         & Yes      & Yes    \\
4FGL J1031.3+5053  & 1ES 1028+511               & 10:31:19 & +50:53:36  & 0.360 & HSP & 0.969 & SDS         & Yes      & Yes    \\
4FGL J1033.5+4221  & GB6 J1033+4222             & 10:33:18 & +42:22:35  & 0.211 & HSP & 0.992 & SDS         & Yes      & Yes    \\
4FGL J1035.6+4409  & 7C 1032+4424               & 10:35:32 & +44:09:31  & 0.444 & LSP & 0.984 & SDS         & No       & No     \\
4FGL J1037.7+5711  & GB6 J1037+5711             & 10:37:44 & +57:11:56  & 1.096 & ISP & 0.807 & SDS         & No       & No     \\
4FGL J1049.7+5011  & NVSS J104857+500943        & 10:48:58 & +50:09:45  & 0.403 & HSP & 0.994 & SDS         & Yes      & Yes    \\
4FGL J1051.9+0103  & NVSS J105151+010312        & 10:51:52 & +01:03:11  & 0.265 & ISP & 0.994 & SDS         & Unc.     & Yes    \\
4FGL J1053.7+4930  & GB6 J1053+4930             & 10:53:44 & +49:29:56  & 0.140 & ISP & 0.987 & SDS         & Yes      & Yes    \\
4FGL J1058.4+0133  & 4C +01.28                  & 10:58:30 & +01:33:59  & 0.894 & LSP & 0.978 & SDS         & Unc.     & No     \\
4FGL J1058.6+5627  & TXS 1055+567               & 10:58:38 & +56:28:11  & 0.143 & ISP & 0.940 & SDS         & Unc.     & Yes    \\
4FGL J1103.6--2329 & 1ES 1101-232               & 11:03:38 & --23:29:31 & 0.186 & HSP & 0.771 & BAS         & No       & No     \\
4FGL J1104.0+0020  & NVSS J110356+002238        & 11:03:56 & +00:22:36  & 0.277 & HSP & 0.994 & SDS         & Yes      & Yes    \\
4FGL J1104.4+3812  & Mkn 421                    & 11:04:27 & +38:12:32  & 0.030 & HSP & 0.754 & BAS         & Yes      & Yes    \\
4FGL J1105.8+3944  & GB6 J1105+3946             & 11:05:54 & +39:46:57  & 0.099 & LSP & 0.986 & SDS         & Yes      & Yes    \\
4FGL J1107.8+1501  & RX J1107.7+1502            & 11:07:48 & +15:02:11  & 0.201 & ISP & 0.989 & SDS         & No       & Yes    \\
4FGL J1109.6+3735  & NVSS J110938+373609        & 11:09:38 & +37:36:12  & 0.397 & ISP & 0.993 & SDS         & Unc.     & Yes    \\
4FGL J1112.4+1751  & 1RXS J111224.2+175131      & 11:12:25 & +17:51:22  & 0.421 & HSP & 0.996 & SDS         & Unc.     & Yes    \\
4FGL J1117.0+2013  & RBS 0958                   & 11:17:06 & +20:14:07  & 0.139 & HSP & 0.971 & SDS         & Yes      & Yes    \\
4FGL J1117.2+0008  & RX J1117.2+0006            & 11:17:18 & +00:06:34  & 0.451 & HSP & 0.994 & SDS         & Unc.     & Yes    \\
4FGL J1120.8+4212  & RBS 0970                   & 11:20:48 & +42:12:12  & 0.568 & HSP & 0.952 & SDS         & No       & Yes    \\
4FGL J1131.4+5809  & 1RXS J113117.8+580911      & 11:31:19 & +58:08:59  & 0.360 & ISP & 0.964 & SDS         & Unc.     & Yes    \\
4FGL J1136.4+6736  & RX J1136.5+6737            & 11:36:30 & +67:37:04  & 0.134 & HSP & 0.989 & SDS         & Yes      & Yes    \\
4FGL J1136.8+2550  & RX J1136.8+2551            & 11:36:50 & +25:50:52  & 0.154 & LSP & 0.979 & SDS         & Yes      & Yes    \\
4FGL J1140.5+1528  & NVSS J114023+152808        & 11:40:23 & +15:28:10  & 0.244 & HSP & 0.994 & SDS         & Yes      & Yes    \\
4FGL J1145.5--0340 & RBS 1029                   & 11:45:35 & --03:40:02 & 0.167 & HSP & 0.990 & SDS         & Yes      & Yes    \\
4FGL J1149.4+2441  & RX J1149.5+2439            & 11:49:30 & +24:39:27  & 0.402 & HSP & 0.994 & SDS         & Unc.     & Yes    \\
\hline
\end{tabular}
\end{table*}
\clearpage
\begin{table*}
\contcaption{}
\begin{tabular}{llcccccccc}
\hline
\hline
Object name        & Counterpart                & R.A.     & Dec.       & $z$   & SED & $S$   & Source & Line Id. & Accept \\
\hline
4FGL J1150.6+4154  & RBS 1040                   & 11:50:35 & +41:54:40  & 1.125 & HSP & 0.803 & SDS         & No       & No     \\
4FGL J1152.1+2837  & GB6 J1152+2837             & 11:52:11 & +28:37:21  & 0.441 & ISP & 0.992 & SDS         & Unc.     & Yes    \\
4FGL J1153.7+3822  & B3 1151+386                & 11:53:43 & +38:23:06  & 0.410 & LSP & 0.993 & SDS         & Yes      & Yes    \\
4FGL J1154.0--0010 & 1RXS J115404.9-001008      & 11:54:05 & --00:10:10 & 0.253 & HSP & 0.993 & SDS         & Yes      & Yes    \\
4FGL J1202.4+4442  & B3 1159+450                & 12:02:09 & +44:44:22  & 0.297 & ISP & 0.994 & SDS         & Yes      & Yes    \\
4FGL J1203.1+6031  & SBS 1200+608               & 12:03:04 & +60:31:19  & 0.065 & LSP & 0.989 & SDS         & Yes      & Yes    \\
4FGL J1203.4--3925 & PMN J1203-3926             & 12:03:18 & --39:26:21 & 0.227 & HSP & 0.786 & ZBL         & Yes      & Yes    \\
4FGL J1208.4+6121  & RGB J1208+613              & 12:08:37 & +61:21:06  & 0.275 & ISP & 0.993 & SDS         & Yes      & Yes    \\
4FGL J1212.0+2242  & RX J1211.9+2242            & 12:11:59 & +22:42:32  & 0.450 & HSP & 0.991 & SDS         & Unc.     & Yes    \\
4FGL J1215.1+0731  & 1ES 1212+078               & 12:15:11 & +07:32:05  & 0.137 & LSP & 0.835 & ZBL         & Yes      & Yes    \\
4FGL J1216.1+0930  & TXS 1213+097               & 12:16:06 & +09:29:10  & 0.094 & HSP & 0.976 & SDS         & Yes      & Yes    \\
4FGL J1217.9+3007  & B2 1215+30                 & 12:17:52 & +30:07:01  & 0.129 & HSP & 0.779 & ZBL         & No       & No     \\
4FGL J1218.0--0028 & PKS 1215-002               & 12:17:59 & --00:29:46 & 0.419 & LSP & 0.991 & SDS         & Unc.     & Yes    \\
4FGL J1219.7--0313 & 1RXS J121946.0-031419      & 12:19:46 & --03:14:24 & 0.299 & ISP & 0.979 & SDS         & Yes      & Yes    \\
4FGL J1221.3+3010  & PG 1218+304                & 12:21:22 & +30:10:37  & 0.184 & HSP & 0.985 & SDS         & Yes      & Yes    \\
4FGL J1221.5+2814  & W Comae                    & 12:21:32 & +28:13:59  & 0.102 & ISP & 0.781 & ZBL         & No       & No     \\
4FGL J1223.8+4649  & RX J1223.8+4651            & 12:23:53 & +46:50:48  & 0.260 & HSP & 0.900 & SDS         & Yes      & Yes    \\
4FGL J1224.4+2436  & MS 1221.8+2452             & 12:24:24 & +24:36:24  & 0.219 & HSP & 0.992 & SDS         & Unc.     & Yes    \\
4FGL J1230.9+3711  & WISE J123124.08+371102.2   & 12:31:24 & +37:11:02  & 0.219 & ISP & 0.989 & ZBL         & Yes      & Yes    \\
4FGL J1231.5+1421  & GB6 J1231+1421             & 12:31:24 & +14:21:24  & 0.256 & ISP & 0.994 & SDS         & Yes      & Yes    \\
4FGL J1231.6+6415  & MS 1229.2+6430             & 12:31:31 & +64:14:18  & 0.163 & ISP & 0.988 & SDS         & Yes      & Yes    \\
4FGL J1233.6+5027  & TXS 1231+507               & 12:33:49 & +50:26:23  & 0.207 & ISP & 0.990 & SDS         & Yes      & Yes    \\
4FGL J1236.3+3858  & RX J1236.4+3859            & 12:36:23 & +39:00:01  & 0.389 & ISP & 0.990 & SDS         & Unc.     & Yes    \\
4FGL J1246.3+0112  & PMN J1246+0113             & 12:46:03 & +01:13:19  & 0.386 & LSP & 0.988 & SDS         & Unc.     & Yes    \\
4FGL J1247.0+4421  & RX J1246.9+4423            & 12:47:01 & +44:23:20  & 0.569 & HSP & 0.890 & ZBL         & No       & No     \\
4FGL J1248.7+5127  & RX J1248.4+5128            & 12:48:34 & +51:28:08  & 0.351 & ISP & 0.991 & SDS         & Yes      & Yes    \\
4FGL J1250.6+0217  & PKS 1247+025               & 12:50:33 & +02:16:32  & 0.955 & LSP & 0.948 & ZBL         & Yes      & No     \\
4FGL J1251.2+1039  & 1RXS J125117.4+103914      & 12:51:18 & +10:39:07  & 0.246 & HSP & 0.990 & SDS         & Yes      & Yes    \\
4FGL J1253.8+0327  & MG1 J125348+0326           & 12:53:47 & +03:26:30  & 0.066 & HSP & 0.988 & SDS         & Yes      & Yes    \\
4FGL J1257.2+3646  & RX J1257.3+3647            & 12:57:17 & +36:47:15  & 0.531 & ISP & 0.983 & SDS         & Unc.     & Yes    \\
4FGL J1257.6+2413  & 1ES 1255+244               & 12:57:32 & +24:12:40  & 0.141 & HSP & 0.978 & SDS         & Yes      & Yes    \\
4FGL J1259.1--2311 & PKS B1256-229              & 12:59:08 & --23:10:39 & 0.481 & LSP & 0.693 & ZBL         & Unc.     & No     \\
4FGL J1304.0+3704  & WISE J130407.31+370908.1   & 13:04:07 & +37:09:08  & 0.940 & LSP & 0.990 & SDS         & No       & Yes    \\
4FGL J1309.4+4305  & B3 1307+433                & 13:09:26 & +43:05:06  & 0.694 & ISP & 0.991 & SDS         & No       & No     \\
4FGL J1319.5+1404  & RX J1319.4+1405            & 13:19:32 & +14:05:33  & 0.573 & ISP & 0.994 & SDS         & Yes      & Yes    \\
4FGL J1322.9+0437  & RBS 1257                   & 13:23:01 & +04:39:51  & 0.224 & HSP & 0.991 & SDS         & Yes      & Yes    \\
4FGL J1326.1+1232  & LEDA 1410672               & 13:26:18 & +12:29:59  & 0.204 & HSP & 0.991 & SDS         & Yes      & Yes    \\
4FGL J1336.2+2320  & 2MASS J13361219+2319581    & 13:36:12 & +23:19:58  & 0.267 & HSP & 0.992 & SDS         & Yes      & Yes    \\
4FGL J1340.5+4409  & RX J1340.4+4410            & 13:40:30 & +44:10:04  & 0.545 & HSP & 0.984 & SDS         & Unc.     & Yes    \\
4FGL J1340.8--0409 & NVSS J134042-041006        & 13:40:42 & --04:10:07 & 0.223 & HSP & 0.698 & ZBL         & Yes      & Yes    \\
4FGL J1341.6+5515  & SBS 1339+554               & 13:41:36 & +55:14:37  & 0.207 & ISP & 0.994 & SDS         & Yes      & Yes    \\
4FGL J1353.3+1434  & OP 186                     & 13:53:23 & +14:35:39  & 0.808 & LSP & 0.994 & SDS         & Unc.     & Yes    \\
4FGL J1353.4+5600  & RX J1353.4+5601            & 13:53:28 & +56:00:57  & 0.404 & ISP & 0.992 & SDS         & Yes      & Yes    \\
4FGL J1354.7+0623  & NVSS J135444+062249        & 13:54:44 & +06:22:48  & 0.276 & ISP & 0.991 & SDS         & Yes      & Yes    \\
4FGL J1402.6+1600  & 4C +16.39                  & 14:02:45 & +15:59:57  & 0.245 & ISP & 0.961 & SDS         & Yes      & Yes    \\
4FGL J1403.4+4319  & NVSS J140319+432018        & 14:03:19 & +43:20:20  & 0.493 & ISP & 0.989 & SDS         & Unc.     & Yes    \\
4FGL J1404.8+6554  & NVSS J140450+655428        & 14:04:50 & +65:54:32  & 0.363 & LSP & 0.955 & SDS         & Unc.     & Yes    \\
4FGL J1417.9+2543  & 1E 1415.6+2557             & 14:17:57 & +25:43:26  & 0.235 & HSP & 0.992 & BAS         & Yes      & Yes    \\
4FGL J1419.3+0444  & 2MASS J14192748+0445138    & 14:19:27 & +04:45:14  & 0.143 & ISP & 0.990 & SDS         & No       & Yes    \\
4FGL J1419.8+5423  & OQ 530                     & 14:19:47 & +54:23:15  & 0.153 & LSP & 0.970 & SDS         & Unc.     & Yes    \\
4FGL J1424.2+0433  & TXS 1421+048               & 14:24:10 & +04:34:52  & 0.666 & LSP & 0.929 & SDS         & Unc.     & No     \\
4FGL J1427.0+2348  & PKS 1424+240               & 14:27:00 & +23:48:00  & 0.604 & HSP & 0.787 & ZBL         & No       & Yes    \\
4FGL J1428.5+4240  & H 1426+428                 & 14:28:33 & +42:40:21  & 0.129 & HSP & 0.959 & SDS         & Yes      & Yes    \\
4FGL J1435.5+2021  & TXS 1433+205               & 14:35:22 & +20:21:18  & 0.748 & LSP & 0.993 & SDS         & No       & Yes    \\
4FGL J1439.9--3953 & 1RXS J143949.8-395524      & 14:39:51 & --39:55:19 & 0.300 & HSP & 0.948 & ZBL         & Yes      & Yes    \\
4FGL J1440.9+0609  & PMN J1440+0610             & 14:40:53 & +06:10:16  & 0.396 & ISP & 0.297 & ZBL         & No       & No     \\
4FGL J1442.7+1200  & 1ES 1440+122               & 14:42:48 & +12:00:40  & 0.163 & HSP & 0.993 & SDS         & Yes      & Yes    \\
4FGL J1501.0+2238  & MS 1458.8+2249             & 15:01:02 & +22:38:06  & 0.235 & ISP & 0.905 & SDS         & No       & No     \\
4FGL J1503.5+4759  & TXS 1501+481               & 15:03:25 & +47:58:30  & 0.344 & LSP & 0.722 & SDS         & Yes      & Yes    \\
4FGL J1506.4+4331  & NVSS J150617+433413        & 15:06:18 & +43:34:14  & 0.470 & ISP & 0.990 & SDS         & Unc.     & Yes    \\
4FGL J1507.2+1721  & NVSS J150716+172103        & 15:07:16 & +17:21:03  & 0.572 & ISP & 0.992 & SDS         & No       & Yes    \\
4FGL J1508.8+2708  & RBS 1467                   & 15:08:43 & +27:09:08  & 0.270 & HSP & 0.990 & SDS         & Yes      & Yes    \\
4FGL J1509.7+5556  & SBS 1508+561               & 15:09:48 & +55:56:17  & 1.398 & ISP & 0.979 & SDS         & No       & Yes    \\
4FGL J1518.6+4044  & GB6 J1518+4045             & 15:18:39 & +40:45:00  & 0.066 & HSP & 0.983 & SDS         & Yes      & Yes    \\
\hline
\end{tabular}
\end{table*}
\clearpage
\begin{table*}
\contcaption{}
\begin{tabular}{llcccccccc}
\hline
\hline
Object name        & Counterpart                & R.A.     & Dec.       & $z$   & SED & $S$   & Source & Line Id. & Accept \\
\hline
4FGL J1522.6--2730 & PKS 1519-273               & 15:22:38 & --27:30:11 & 1.297 & LSP & 0.918 & ZBL         & No       & Yes    \\
4FGL J1523.2+0533  & NVSS J152312+053357        & 15:23:13 & +05:33:55  & 0.176 & LSP & 0.989 & SDS         & Yes      & Yes    \\
4FGL J1532.0+3016  & RX J1531.9+3016            & 15:32:02 & +30:16:29  & 0.065 & ISP & 0.991 & SDS         & Yes      & Yes    \\
4FGL J1533.2+1855  & RX J1533.1+1854            & 15:33:11 & +18:54:29  & 0.307 & ISP & 0.995 & SDS         & Yes      & Yes    \\
4FGL J1533.2+3416  & RX J1533.3+3416            & 15:33:24 & +34:16:40  & 0.329 & HSP & 0.981 & SDS         & No       & No     \\
4FGL J1534.8+3716  & RGB J1534+372              & 15:34:47 & +37:15:55  & 0.143 & ISP & 0.994 & SDS         & Yes      & Yes    \\
4FGL J1535.0+5320  & 1ES 1533+535               & 15:35:01 & +53:20:37  & 0.875 & HSP & 0.977 & SDS         & No       & No     \\
4FGL J1535.4+3919  & RX J1535.4+3922            & 15:35:29 & +39:22:46  & 0.257 & ISP & 0.980 & SDS         & Yes      & Yes    \\
4FGL J1540.7+1449  & 4C +14.60                  & 15:40:49 & +14:47:46  & 0.606 & LSP & 0.984 & SDS         & Unc.     & Yes    \\
4FGL J1541.7+1413  & WISE J154150.09+141437.6   & 15:41:50 & +14:14:38  & 0.223 & LSP & 0.992 & SDS         & Yes      & Yes    \\
4FGL J1548.3+1456  & NVSS J154824+145702        & 15:48:24 & +14:57:03  & 0.230 & ISP & 0.473 & ZBL         & Yes      & Yes    \\
4FGL J1552.0+0850  & TXS 1549+089               & 15:52:03 & +08:50:47  & 0.125 & LSP & 0.977 & SDS         & No       & No     \\
4FGL J1606.2+1346  & MG1 J160619+1345           & 16:06:18 & +13:45:33  & 0.290 & HSP & 0.989 & SDS         & Yes      & Yes    \\
4FGL J1616.7+4107  & B3 1615+412                & 16:17:06 & +41:06:47  & 0.267 & LSP & 0.989 & SDS         & Yes      & Yes    \\
4FGL J1626.3+3514  & RGB J1626+352              & 16:26:26 & +35:13:41  & 0.498 & HSP & 0.986 & SDS         & Yes      & Yes    \\
4FGL J1626.6--7639 & PKS 1619-765               & 16:26:38 & --76:38:56 & 0.105 & ISP & 0.867 & ZBL         & Yes      & Yes    \\
4FGL J1637.1+1316  & 1RXS J163717.1+131418      & 16:37:17 & +13:14:39  & 0.655 & ISP & 0.910 & ZBL         & Unc.     & Yes    \\
4FGL J1637.2+4327  & 1RXS J163711.1+432548      & 16:37:10 & +43:26:00  & 0.343 & HSP & 0.992 & SDS         & Unc.     & Yes    \\
4FGL J1637.6+4548  & B3 1635+458                & 16:37:27 & +45:47:49  & 0.192 & HSP & 0.994 & SDS         & Yes      & Yes    \\
4FGL J1642.4+2211  & 1RXS J164220.4+221132      & 16:42:20 & +22:11:43  & 0.592 & HSP & 0.993 & SDS         & Unc.     & Yes    \\
4FGL J1643.0+3223  & NVSS J164301+322104        & 16:43:01 & +32:21:04  & 0.371 & LSP & 0.993 & SDS         & No       & Yes    \\
4FGL J1647.5+2911  & B2 1645+29                 & 16:47:27 & +29:09:50  & 0.133 & ISP & 0.986 & SDS         & Yes      & Yes    \\
4FGL J1652.7+4024  & RX J1652.7+4023            & 16:52:50 & +40:23:10  & 1.803 & HSP & 0.991 & SDS         & No       & Yes    \\
4FGL J1653.8+3945  & Mkn 501                    & 16:53:52 & +39:45:37  & 0.033 & HSP & 0.878 & BAS         & Yes      & Yes    \\
4FGL J1704.2+1234  & NVSS J170409+123421        & 17:04:10 & +12:34:22  & 0.452 & LSP & 0.749 & ZBL         & Yes      & No     \\
4FGL J1706.8+3004  & 87GB 170454.3+300758       & 17:06:50 & +30:04:13  & 1.456 & ISP & 0.989 & SDS         & No       & Yes    \\
4FGL J1725.0+1152  & 1H 1720+117                & 17:25:04 & +11:52:15  & 0.018 & HSP & 0.910 & ASG         & No       & No     \\
4FGL J1733.4+5428  & SDSS J173340.31+542636.9   & 17:33:40 & +54:26:37  & 0.417 & ISP & 0.996 & SDS         & Unc.     & Yes    \\
4FGL J1800.6+7828  & S5 1803+784                & 18:00:46 & +78:28:04  & 0.683 & LSP & 0.674 & ZBL         & Yes      & No     \\
4FGL J1917.7--1921 & 1H 1914-194                & 19:17:45 & --19:21:32 & 0.137 & HSP & 0.449 & ZBL         & Unc.     & No     \\
4FGL J1954.9--5640 & 1RXS J195503.1-564031      & 19:55:03 & --56:40:29 & 0.221 & HSP & 0.916 & ZBL         & Yes      & Yes    \\
4FGL J2000.0+6508  & 1ES 1959+650               & 20:00:00 & +65:08:55  & 0.048 & HSP & 0.992 & ASG         & No       & Yes    \\
4FGL J2009.4--4849 & PKS 2005-489               & 20:09:25 & --48:49:54 & 0.071 & HSP & 0.400 & ZBL         & No       & No     \\
4FGL J2054.8+0015  & RGB J2054+002              & 20:54:57 & +00:15:38  & 0.151 & LSP & 0.992 & SDS         & Yes      & Yes    \\
4FGL J2055.4--0504 & NVSS J205523-050618        & 20:55:23 & --05:06:19 & 0.342 & ISP & 0.988 & SDS         & Yes      & Yes    \\
4FGL J2115.9--0113 & NVSS J211603-010828        & 21:16:03 & --01:08:28 & 0.305 & LSP & 0.992 & SDS         & Yes      & Yes    \\
4FGL J2131.5--0916 & RBS 1752                   & 21:31:35 & --09:15:24 & 0.448 & HSP & 0.851 & ZBL         & Unc.     & Yes    \\
4FGL J2150.7--1750 & MRSS 600-040574            & 21:50:47 & --17:49:54 & 0.185 & ISP & 0.302 & ZBL         & Yes      & No     \\
4FGL J2153.1--0041 & RBS 1792                   & 21:53:05 & --00:42:31 & 0.342 & ISP & 0.991 & SDS         & Unc.     & Yes    \\
4FGL J2158.8--3013 & PKS 2155-304               & 21:58:52 & --30:13:32 & 0.117 & HSP & 0.760 & OTE         & No       & Yes    \\
4FGL J2159.1--2840 & LEDA 3218689               & 21:59:11 & --28:41:16 & 0.270 & HSP & 0.908 & ZBL         & Yes      & Yes    \\
4FGL J2202.7+4216  & BL Lac                     & 22:02:43 & +42:16:40  & 0.069 & LSP & 0.758 & OTE         & No       & Yes    \\
4FGL J2204.3+0438  & 4C +04.77                  & 22:04:18 & +04:40:02  & 0.027 & ISP & 0.860 & ZBL         & Yes      & No     \\
4FGL J2206.8--0032 & PMN J2206-0031             & 22:06:43 & --00:31:02 & 1.053 & LSP & 0.993 & SDS         & No       & Yes    \\
4FGL J2211.0--0003 & RX J2211.1-0003            & 22:11:08 & --00:03:03 & 0.362 & ISP & 0.980 & SDS         & Unc.     & Yes    \\
4FGL J2220.5+2813  & RX J2220.4+2814            & 22:20:29 & +28:13:56  & 0.149 & HSP & 0.982 & SDS         & Yes      & Yes    \\
4FGL J2225.5--1114 & PKS 2223-114               & 22:25:44 & --11:13:41 & 0.997 & LSP & 0.963 & ZBL         & No       & Yes    \\
4FGL J2228.6--1636 & 2MASS J22283018-1636432    & 22:28:30 & --16:36:43 & 0.525 & ISP & 0.896 & ZBL         & Unc.     & Yes    \\
4FGL J2232.8+1334  & RX J2233.0+1335            & 22:33:01 & +13:36:02  & 0.214 & HSP & 0.994 & SDS         & Yes      & Yes    \\
4FGL J2244.6+2502  & NVSS J224436+250345        & 22:44:37 & +25:03:43  & 0.650 & ISP & 0.776 & ZBL         & No       & No     \\
4FGL J2244.9--0007 & NVSS J224448-000616        & 22:44:48 & --00:06:20 & 0.640 & HSP & 0.920 & ZBL         & Unc.     & Yes    \\
4FGL J2245.9+1544  & 87GB 224338.7+152914       & 22:46:05 & +15:44:35  & 0.596 & ISP & 0.876 & ZBL         & Unc.     & Yes    \\
4FGL J2247.4--0001 & PKS 2244-002               & 22:47:30 & +00:00:06  & 0.108 & LSP & 0.991 & SDS         & No       & Yes    \\
4FGL J2251.7--3208 & 1RXS J225146.9-320614      & 22:51:48 & --32:06:13 & 0.246 & HSP & 0.957 & BAS         & No       & No     \\
4FGL J2252.6+1245  & 2MASS J22523220+1245109    & 22:52:32 & +12:45:11  & 0.497 & ISP & 0.992 & SDS         & Unc.     & Yes    \\
4FGL J2255.2+2411  & MG3 J225517+2409           & 22:55:15 & +24:10:11  & 1.370 & LSP & 0.986 & SDS         & No       & Yes    \\
4FGL J2314.0+1445  & RGB J2313+147              & 23:13:57 & +14:44:23  & 0.164 & HSP & 0.981 & SDS         & Yes      & Yes    \\
4FGL J2343.6+3438  & 1RXS J234332.5+343957      & 23:43:34 & +34:39:51  & 0.366 & HSP & 0.990 & SDS         & Yes      & Yes    \\
4FGL J2354.1+2720  & NVSS J235402+272328        & 23:54:02 & +27:23:28  & 0.149 & LSP & 0.991 & SDS         & No       & Yes    \\
4FGL J2357.4--0152 & PKS 2354-021               & 23:57:25 & --01:52:16 & 0.812 & LSP & 0.971 & ZBL         & Unc.     & No     \\
4FGL J2359.0--3038 & H 2356-309                 & 23:59:08 & --30:37:41 & 0.165 & HSP & 0.892 & BAS         & No       & No     \\
\hline
\end{tabular}
\end{table*}

\clearpage
\begin{landscape}
\begin{footnotesize}
\begin{table}
\caption{Results of the multi-component fits and spectral index measurements for objects with valid fits. The table columns report, respectively, the source name,the synchrotron peak frequency, the logarithm of the rest frame $\nu L_\nu$ at $5500\,$\AA, the power-law spectral index in $\gamma$ rays, the frequency of normalization, the normalization of the host component, the normalization of the optical power-law component, the optical spectral index $\alpha_{opt}$, the spectral index difference, the ratio between the synchrotron and host galaxy specific flux at rest frame wavelength $\lambda = 5500\,$\AA\ and integrated ratio computed for $3500\, \mathrm{\AA} \leqslant \lambda \leqslant 8000\,$\AA. The errors on the spectral decompositions correspond to statistical uncertainties only. \label{tab02}}
\begin{center}
\begin{tabular}{lcccccccccc}
\hline
\hline
Object name        & $\log \nu_{Peak}^{Syn}$ & $\log \nu L_\nu$                & $\alpha_\gamma$ & $\nu_\mathrm{norm}$ & Host norm.         & PL norm. & $\alpha_{opt}$ & $\alpha_\gamma - \alpha_{opt}$ & $f_{ratio}$ & $f_{ratio}^{int}$ \\
                   & Hz                      & $\mathrm{erg\, s^{-1}}$ &               & $10^{14}\,$Hz      & $10^{-4}\,$Jy       & $10^{-4}\,$Jy    &               &                             &           &                \\
\hline
4FGL J0003.2+2207  & 13.920 & $43.7 \pm  0.3$ & $ 2.21 \pm  0.21$ & 4.996 & $ 2.06 \pm  0.33$ & $  0.83 \pm 0.27$ & $+2.32 \pm 0.20$ & $ -0.11 \pm  0.40$ & $   0.47 \pm      0.23$ & $   0.38 \pm      0.18$ \\
4FGL J0017.8+1455  & 14.200 & $45.1 \pm  1.5$ & $ 2.19 \pm  0.11$ & 4.996 & $ 0.91 \pm  0.97$ & $  1.77 \pm 0.42$ & $+0.99 \pm 0.61$ & $ +1.20 \pm  0.73$ & $   2.67 \pm      3.46$ & $   2.04 \pm      2.65$ \\
4FGL J0022.0+0006  & 17.130 & $44.7 \pm  3.3$ & $ 1.53 \pm  0.16$ & 4.996 & $ 1.07 \pm  1.03$ & $  0.22 \pm 0.44$ & $+0.99 \pm 5.14$ & $ +0.54 \pm  5.30$ & $   0.29 \pm      0.84$ & $   0.22 \pm      0.64$ \\
4FGL J0028.8--0112 & 14.000 & $44.2 \pm  0.3$ & $ 2.15 \pm  0.23$ & 4.996 & $11.52 \pm  1.29$ & $  1.02 \pm 1.10$ & $+1.39 \pm 1.46$ & $ +0.75 \pm  1.69$ & $   0.10 \pm      0.12$ & $   0.09 \pm      0.11$ \\
4FGL J0035.2+1514  & 14.987 & $47.9 \pm  3.9$ & $ 1.85 \pm  0.04$ & 4.995 & $ 3.28 \pm  2.70$ & $  9.31 \pm 0.12$ & $+0.50 \pm 0.05$ & $ +1.36 \pm  0.09$ & $   4.51 \pm      3.77$ & $   3.20 \pm      2.67$ \\
4FGL J0040.4--2340 & 14.500 & $45.0 \pm  0.2$ & $ 2.35 \pm  0.13$ & 4.996 & $ 3.41 \pm  0.77$ & $  4.12 \pm 0.10$ & $+1.49 \pm 0.40$ & $ +0.86 \pm  0.53$ & $   0.90 \pm      0.38$ & $   0.66 \pm      0.28$ \\
4FGL J0056.3--0935 & 14.165 & $44.3 \pm  0.5$ & $ 1.87 \pm  0.05$ & 4.996 & $ 6.60 \pm  1.85$ & $  3.15 \pm 1.57$ & $+0.74 \pm 1.00$ & $ +1.14 \pm  1.05$ & $   0.56 \pm      0.43$ & $   0.52 \pm      0.41$ \\
4FGL J0059.3--0152 & 16.965 & $44.3 \pm  0.5$ & $ 1.81 \pm  0.11$ & 4.996 & $ 2.45 \pm  0.75$ & $  2.19 \pm 0.50$ & $+1.84 \pm 0.23$ & $ -0.04 \pm  0.34$ & $   1.12 \pm      0.60$ & $   0.86 \pm      0.46$ \\
4FGL J0101.0--0059 & 14.200 & $47.2 \pm  1.2$ & $ 2.38 \pm  0.18$ & 4.995 & undetected        & $  1.40 \pm 0.07$ & $+2.05 \pm 0.13$ & $ +0.33 \pm  0.31$ & --                      & --                      \\
4FGL J0111.4+0534  & 14.480 & $45.1 \pm  1.3$ & $ 1.92 \pm  0.19$ & 4.996 & $ 1.37 \pm  0.52$ & $  0.54 \pm 0.24$ & $+0.21 \pm 1.28$ & $ +1.71 \pm  1.47$ & $   0.47 \pm      0.39$ & $   0.46 \pm      0.38$ \\
4FGL J0115.8+2519  & 15.752 & $45.6 \pm  3.1$ & $ 1.92 \pm  0.03$ & 4.996 & $ 1.11 \pm  3.29$ & $  3.80 \pm 1.11$ & $+1.11 \pm 0.76$ & $ +0.81 \pm  0.79$ & $   5.12 \pm     16.72$ & $   3.55 \pm     11.60$ \\
4FGL J0121.8--3916 & 15.200 & $45.7 \pm  0.5$ & $ 1.90 \pm  0.09$ & 4.996 & $ 3.27 \pm  0.49$ & $  1.73 \pm 0.15$ & $-0.38 \pm 0.23$ & $ +2.27 \pm  0.33$ & $   0.56 \pm      0.13$ & $   0.70 \pm      0.17$ \\
4FGL J0152.6+0147  & 16.580 & $44.2 \pm  0.7$ & $ 1.98 \pm  0.05$ & 4.995 & $12.71 \pm  4.28$ & $  4.28 \pm 3.88$ & $+1.06 \pm 1.60$ & $ +0.93 \pm  1.66$ & $   0.38 \pm      0.48$ & $   0.35 \pm      0.44$ \\
4FGL J0201.1+0036  & 16.910 & $44.9 \pm  1.6$ & $ 1.88 \pm  0.20$ & 4.995 & $ 0.97 \pm  0.83$ & $  1.14 \pm 0.38$ & $+0.73 \pm 0.90$ & $ +1.15 \pm  1.10$ & $   1.54 \pm      1.82$ & $   1.28 \pm      1.51$ \\
4FGL J0220.8--0841 & 15.150 & $46.0 \pm  0.9$ & $ 1.74 \pm  0.16$ & 4.995 & $ 1.46 \pm  0.50$ & $  2.15 \pm 0.14$ & $+0.06 \pm 0.22$ & $ +1.68 \pm  0.37$ & $   1.74 \pm      0.71$ & $   1.79 \pm      0.73$ \\
4FGL J0237.6--3602 & 16.030 & $45.9 \pm  0.2$ & $ 1.86 \pm  0.09$ & 4.996 & $ 0.95 \pm  0.28$ & $  4.97 \pm 0.10$ & $+0.45 \pm 0.06$ & $ +1.42 \pm  0.15$ & $   6.75 \pm      2.12$ & $   5.92 \pm      1.86$ \\
4FGL J0238.7+2555  & 15.500 & $45.8 \pm  1.2$ & $ 1.94 \pm  0.17$ & 4.996 & $ 0.78 \pm  0.23$ & $  0.67 \pm 0.04$ & $+1.05 \pm 0.22$ & $ +0.89 \pm  0.38$ & $   1.46 \pm      0.52$ & $   0.89 \pm      0.32$ \\
4FGL J0242.9+0045  & 14.765 & $45.1 \pm  3.6$ & $ 2.10 \pm  0.17$ & 4.996 & $ 0.78 \pm  0.84$ & $  0.40 \pm 0.25$ & $+1.25 \pm 1.79$ & $ +0.86 \pm  1.96$ & $   0.82 \pm      1.38$ & $   0.52 \pm      0.88$ \\
4FGL J0250.6+1712  & 17.060 & $45.3 \pm  0.3$ & $ 1.84 \pm  0.08$ & 4.996 & $ 3.18 \pm  1.01$ & $  6.37 \pm 0.08$ & $+0.80 \pm 0.31$ & $ +1.04 \pm  0.39$ & $   1.90 \pm      0.83$ & $   1.60 \pm      0.70$ \\
4FGL J0301.0--1652 & 13.600 & $44.9 \pm  0.6$ & $ 2.31 \pm  0.17$ & 4.996 & $ 1.18 \pm  0.39$ & $  1.43 \pm 0.18$ & $+1.35 \pm 0.31$ & $ +0.97 \pm  0.48$ & $   1.73 \pm      0.79$ & $   1.22 \pm      0.56$ \\
4FGL J0304.5--0054 & 15.350 & $46.0 \pm  1.1$ & $ 2.05 \pm  0.13$ & 4.995 & $ 1.16 \pm  0.72$ & $  2.18 \pm 0.05$ & $+0.65 \pm 0.32$ & $ +1.40 \pm  0.45$ & $   2.36 \pm      1.69$ & $   1.82 \pm      1.30$ \\
4FGL J0305.1--1608 & 16.415 & $45.2 \pm  0.5$ & $ 1.79 \pm  0.10$ & 4.996 & $ 1.30 \pm  0.54$ & $  1.86 \pm 0.04$ & $+0.89 \pm 0.47$ & $ +0.89 \pm  0.57$ & $   1.40 \pm      0.83$ & $   1.10 \pm      0.65$ \\
4FGL J0316.2--2608 & 16.085 & $45.7 \pm  0.3$ & $ 1.87 \pm  0.06$ & 4.996 & $ 0.85 \pm  0.19$ & $  2.28 \pm 0.06$ & $+0.56 \pm 0.07$ & $ +1.31 \pm  0.13$ & $   3.61 \pm      0.89$ & $   2.98 \pm      0.74$ \\
4FGL J0326.2+0225  & 15.904 & $44.4 \pm  1.1$ & $ 1.88 \pm  0.06$ & 4.774 & $ 1.65 \pm  1.88$ & $  3.65 \pm 1.67$ & $+0.46 \pm 0.92$ & $ +1.42 \pm  0.97$ & $   2.56 \pm      4.07$ & $   2.45 \pm      3.90$ \\
4FGL J0340.5--2118 & 13.760 & $44.7 \pm  0.0$ & $ 2.26 \pm  0.04$ & 4.996 & undetected        & $  3.24 \pm 0.01$ & $+1.79 \pm 0.02$ & $ +0.47 \pm  0.06$ & --                      & --                      \\
4FGL J0353.0--6831 & 17.780 & $43.2 \pm  0.9$ & $ 1.67 \pm  0.09$ & 4.996 & $ 1.03 \pm  0.40$ & $  0.20 \pm 0.36$ & $+0.85 \pm 3.21$ & $ +0.82 \pm  3.29$ & $   0.23 \pm      0.49$ & $   0.21 \pm      0.45$ \\
4FGL J0543.9--5531 & 16.690 & $45.6 \pm  0.0$ & $ 1.76 \pm  0.03$ & 4.996 & undetected        & $ 11.80 \pm 0.06$ & $+0.54 \pm 0.02$ & $ +1.23 \pm  0.04$ & --                      & --                      \\
4FGL J0558.0--3837 & 16.580 & $45.7 \pm  0.2$ & $ 1.91 \pm  0.06$ & 4.996 & $ 2.80 \pm  0.57$ & $  7.75 \pm 0.27$ & $+0.43 \pm 0.09$ & $ +1.47 \pm  0.15$ & $   3.44 \pm      0.83$ & $   3.15 \pm      0.75$ \\
4FGL J0710.4+5908  & 17.720 & $44.4 \pm  1.7$ & $ 1.69 \pm  0.05$ & 4.995 & $ 4.99 \pm  4.80$ & $  3.19 \pm 3.93$ & $+0.59 \pm 2.58$ & $ +1.10 \pm  2.63$ & $   0.75 \pm      1.65$ & $   0.71 \pm      1.56$ \\
4FGL J0727.1+3734  & 14.345 & $46.2 \pm  4.1$ & $ 1.83 \pm  0.21$ & 4.996 & $ 0.68 \pm  0.41$ & $  0.42 \pm 0.04$ & $+0.72 \pm 0.46$ & $ +1.11 \pm  0.68$ & $   1.01 \pm      0.70$ & $   0.67 \pm      0.46$ \\
4FGL J0731.9+2805  & 14.750 & $44.9 \pm  1.2$ & $ 2.02 \pm  0.16$ & 4.995 & $ 1.54 \pm  1.11$ & $  1.77 \pm 0.60$ & $+0.72 \pm 0.82$ & $ +1.31 \pm  0.98$ & $   1.46 \pm      1.55$ & $   1.25 \pm      1.33$ \\
4FGL J0733.7+4110  & 14.405 & $44.4 \pm  1.0$ & $ 2.00 \pm  0.08$ & 4.996 & $ 0.40 \pm  0.73$ & $  1.95 \pm 0.44$ & $+1.11 \pm 0.47$ & $ +0.89 \pm  0.55$ & $   6.19 \pm     12.67$ & $   5.02 \pm     10.27$ \\
4FGL J0740.9+3203  & 15.320 & $44.5 \pm  0.5$ & $ 2.44 \pm  0.13$ & 4.995 & $ 2.74 \pm  0.59$ & $  1.20 \pm 0.36$ & $+2.36 \pm 0.27$ & $ +0.08 \pm  0.40$ & $   0.61 \pm      0.31$ & $   0.41 \pm      0.21$ \\
4FGL J0749.2+2314  & 15.300 & $44.3 \pm  0.4$ & $ 1.94 \pm  0.17$ & 4.995 & $ 1.90 \pm  0.40$ & $  0.90 \pm 0.25$ & $+1.78 \pm 0.31$ & $ +0.16 \pm  0.48$ & $   0.63 \pm      0.30$ & $   0.46 \pm      0.22$ \\
4FGL J0758.9+2703  & 13.760 & $43.8 \pm  0.4$ & $ 2.15 \pm  0.07$ & 4.995 & $ 1.57 \pm  0.55$ & $  2.06 \pm 0.44$ & $+1.78 \pm 0.23$ & $ +0.38 \pm  0.30$ & $   1.53 \pm      0.87$ & $   1.27 \pm      0.72$ \\
4FGL J0809.6+3455  & 15.590 & $43.9 \pm  0.2$ & $ 1.80 \pm  0.09$ & 4.996 & $ 4.62 \pm  0.68$ & $  3.11 \pm 0.58$ & $+1.59 \pm 0.22$ & $ +0.21 \pm  0.31$ & $   0.77 \pm      0.26$ & $   0.66 \pm      0.22$ \\
4FGL J0812.9+5555  & 15.205 & $45.2 \pm  1.2$ & $ 2.18 \pm  0.13$ & 4.995 & $ 1.17 \pm  0.34$ & $  0.55 \pm 0.21$ & $-1.50 \pm 1.06$ & $ +3.67 \pm  1.19$ & $   0.38 \pm      0.26$ & $   0.80 \pm      0.54$ \\
4FGL J0818.4+2816  & 14.600 & $44.8 \pm  0.8$ & $ 2.06 \pm  0.11$ & 4.995 & $ 0.82 \pm  0.80$ & $  2.79 \pm 0.47$ & $+0.69 \pm 0.39$ & $ +1.37 \pm  0.50$ & $   4.25 \pm      4.86$ & $   3.72 \pm      4.25$ \\
4FGL J0819.4+4035  & 13.580 & $45.2 \pm  3.4$ & $ 2.00 \pm  0.17$ & 4.996 & $ 0.86 \pm  1.07$ & $  0.68 \pm 0.33$ & $+1.33 \pm 1.35$ & $ +0.67 \pm  1.52$ & $   1.25 \pm      2.18$ & $   0.80 \pm      1.38$ \\
4FGL J0837.3+1458  & 16.820 & $45.4 \pm  0.3$ & $ 1.85 \pm  0.16$ & 4.996 & $ 2.50 \pm  0.76$ & $  4.64 \pm 0.08$ & $+0.11 \pm 0.35$ & $ +1.74 \pm  0.51$ & $   1.59 \pm      0.68$ & $   1.63 \pm      0.69$ \\
4FGL J0842.5+0251  & 14.165 & $45.4 \pm  1.9$ & $ 2.03 \pm  0.17$ & 4.996 & $ 1.11 \pm  0.68$ & $  0.81 \pm 0.22$ & $+0.56 \pm 0.86$ & $ +1.47 \pm  1.02$ & $   0.98 \pm      0.86$ & $   0.82 \pm      0.71$ \\
4FGL J0847.2+1134  & 14.490 & $44.6 \pm  2.1$ & $ 1.72 \pm  0.07$ & 4.996 & $ 1.19 \pm  2.39$ & $  2.51 \pm 1.47$ & $+0.84 \pm 1.25$ & $ +0.89 \pm  1.32$ & $   2.63 \pm      6.84$ & $   2.26 \pm      5.88$ \\
\hline
\end{tabular}
\end{center}
\end{table}
\clearpage
\begin{table}
\contcaption{}
\begin{center}
\begin{tabular}{lcccccccccc}
\hline
\hline
Object name        & $\log \nu_{Peak}^{Syn}$ & $\log \nu L_\nu$                & $\alpha_\gamma$ & $\nu_\mathrm{norm}$ & Host norm.         & PL norm. & $\alpha_{opt}$ & $\alpha_\gamma - \alpha_{opt}$ & $f_{ratio}$ & $f_{ratio}^{int}$ \\
                   & Hz                      & $\mathrm{erg\, s^{-1}}$ &               & $10^{14}\,$Hz      & $10^{-4}\,$Jy       & $10^{-4}\,$Jy    &               &                             &           &                \\
\hline
4FGL J0850.5+3455  & 15.480 & $44.5 \pm  0.5$ & $ 2.01 \pm  0.09$ & 4.995 & $ 3.92 \pm  1.01$ & $  2.25 \pm 0.73$ & $+1.00 \pm 0.63$ & $ +1.01 \pm  0.72$ & $   0.69 \pm      0.40$ & $   0.60 \pm      0.35$ \\
4FGL J0854.0+2753  & 14.120 & $45.5 \pm  4.0$ & $ 1.48 \pm  0.22$ & 4.996 & $ 1.02 \pm  0.89$ & $  0.33 \pm 0.20$ & $+1.19 \pm 2.03$ & $ +0.30 \pm  2.25$ & $   0.55 \pm      0.82$ & $   0.34 \pm      0.50$ \\
4FGL J0857.7+0137  & 15.425 & $45.1 \pm  1.1$ & $ 2.36 \pm  0.15$ & 4.995 & $ 2.45 \pm  1.20$ & $  1.82 \pm 0.55$ & $+1.02 \pm 0.75$ & $ +1.35 \pm  0.91$ & $   1.01 \pm      0.80$ & $   0.78 \pm      0.62$ \\
4FGL J0901.4+4542  & 14.310 & $44.8 \pm  2.0$ & $ 2.27 \pm  0.21$ & 4.996 & $ 1.06 \pm  0.88$ & $  0.71 \pm 0.40$ & $+0.94 \pm 1.44$ & $ +1.32 \pm  1.66$ & $   0.90 \pm      1.26$ & $   0.71 \pm      0.98$ \\
4FGL J0910.8+3859  & 13.850 & $44.6 \pm  1.4$ & $ 2.00 \pm  0.08$ & 4.995 & $ 0.59 \pm  1.38$ & $  2.64 \pm 0.79$ & $+1.37 \pm 0.49$ & $ +0.63 \pm  0.56$ & $   5.84 \pm     15.28$ & $   4.48 \pm     11.71$ \\
4FGL J0911.7+3349  & 15.450 & $45.5 \pm  3.1$ & $ 2.20 \pm  0.11$ & 4.996 & $ 1.12 \pm  1.14$ & $  0.94 \pm 0.28$ & $+1.39 \pm 0.90$ & $ +0.81 \pm  1.02$ & $   1.45 \pm      1.89$ & $   0.84 \pm      1.11$ \\
4FGL J0916.7+5238  & 16.305 & $44.7 \pm  0.7$ & $ 1.84 \pm  0.12$ & 4.996 & $ 3.10 \pm  0.98$ & $  1.38 \pm 0.64$ & $+0.63 \pm 1.10$ & $ +1.21 \pm  1.22$ & $   0.54 \pm      0.42$ & $   0.49 \pm      0.38$ \\
4FGL J0917.3--0342 & 15.810 & $45.1 \pm  2.7$ & $ 1.77 \pm  0.17$ & 4.996 & $ 1.97 \pm  2.05$ & $  1.16 \pm 0.85$ & $+1.12 \pm 2.04$ & $ +0.65 \pm  2.22$ & $   0.84 \pm      1.47$ & $   0.61 \pm      1.08$ \\
4FGL J0930.5+4951  & 17.300 & $44.5 \pm  0.9$ & $ 1.82 \pm  0.12$ & 4.996 & $ 1.48 \pm  0.90$ & $  1.77 \pm 0.58$ & $+0.78 \pm 0.74$ & $ +1.04 \pm  0.87$ & $   1.48 \pm      1.38$ & $   1.29 \pm      1.21$ \\
4FGL J0937.9--1434 & 14.765 & $45.0 \pm  0.8$ & $ 1.98 \pm  0.10$ & 4.996 & $ 1.12 \pm  0.61$ & $  1.96 \pm 0.27$ & $+1.18 \pm 0.34$ & $ +0.80 \pm  0.44$ & $   2.45 \pm      1.66$ & $   1.79 \pm      1.21$ \\
4FGL J0940.4+6148  & 16.470 & $44.6 \pm  0.9$ & $ 1.83 \pm  0.09$ & 4.996 & $ 1.83 \pm  0.71$ & $  0.86 \pm 0.41$ & $+1.06 \pm 1.03$ & $ +0.78 \pm  1.13$ & $   0.61 \pm      0.53$ & $   0.49 \pm      0.43$ \\
4FGL J0942.3+2842  & 15.205 & $44.8 \pm  3.5$ & $ 1.90 \pm  0.16$ & 4.996 & $ 0.61 \pm  0.66$ & $  0.29 \pm 0.23$ & $+1.08 \pm 2.08$ & $ +0.81 \pm  2.24$ & $   0.69 \pm      1.31$ & $   0.49 \pm      0.93$ \\
4FGL J0945.7+5759  & 13.985 & $45.0 \pm  0.6$ & $ 2.09 \pm  0.08$ & 4.995 & $ 2.56 \pm  0.90$ & $  2.79 \pm 0.51$ & $+0.77 \pm 0.44$ & $ +1.33 \pm  0.52$ & $   1.38 \pm      0.73$ & $   1.18 \pm      0.63$ \\
4FGL J0952.8+0712  & 14.290 & $45.8 \pm  1.4$ & $ 1.99 \pm  0.12$ & 4.996 & $ 0.56 \pm  0.31$ & $  1.04 \pm 0.05$ & $+1.09 \pm 0.19$ & $ +0.90 \pm  0.31$ & $   3.18 \pm      1.91$ & $   1.93 \pm      1.15$ \\
4FGL J0955.1+3551  & 14.705 & $46.0 \pm 24.6$ & $ 1.89 \pm  0.17$ & 4.996 & $ 0.16 \pm  1.20$ & $  0.41 \pm 0.04$ & $+1.77 \pm 0.93$ & $ +0.13 \pm  1.11$ & $   7.13 \pm     53.95$ & $   2.39 \pm     18.12$ \\
4FGL J0959.4+2120  & 14.825 & $45.3 \pm  1.5$ & $ 2.18 \pm  0.13$ & 4.995 & $ 1.31 \pm  0.78$ & $  1.12 \pm 0.29$ & $+0.80 \pm 0.72$ & $ +1.38 \pm  0.85$ & $   1.17 \pm      1.01$ & $   0.92 \pm      0.79$ \\
4FGL J1003.6+2605  & 13.285 & $46.4 \pm 20.8$ & $ 2.28 \pm  0.11$ & 4.996 & $ 0.63 \pm  1.21$ & $  0.22 \pm 0.08$ & $+1.32 \pm 1.60$ & $ +0.96 \pm  1.71$ & $   0.87 \pm      1.99$ & $   0.36 \pm      0.82$ \\
4FGL J1012.3+0629  & 13.840 & $46.5 \pm  4.0$ & $ 2.09 \pm  0.05$ & 4.995 & $ 1.02 \pm  1.13$ & $  1.75 \pm 0.13$ & $+1.25 \pm 0.29$ & $ +0.85 \pm  0.34$ & $   3.53 \pm      4.18$ & $   1.76 \pm      2.08$ \\
4FGL J1021.9+5123  & 14.075 & $44.0 \pm  0.7$ & $ 2.09 \pm  0.20$ & 4.996 & $ 1.83 \pm  0.54$ & $  0.44 \pm 0.38$ & $+1.41 \pm 1.36$ & $ +0.68 \pm  1.57$ & $   0.29 \pm      0.34$ & $   0.24 \pm      0.28$ \\
4FGL J1023.8+3002  & 15.500 & $45.5 \pm  2.2$ & $ 1.86 \pm  0.15$ & 4.995 & $ 0.70 \pm  0.80$ & $  1.25 \pm 0.23$ & $+0.89 \pm 0.59$ & $ +0.97 \pm  0.75$ & $   2.64 \pm      3.52$ & $   1.91 \pm      2.55$ \\
4FGL J1024.8+2332  & 13.650 & $44.6 \pm  0.6$ & $ 2.44 \pm  0.09$ & 4.996 & $ 1.74 \pm  1.02$ & $  3.53 \pm 0.65$ & $+1.34 \pm 0.29$ & $ +1.10 \pm  0.38$ & $   2.56 \pm      1.98$ & $   2.05 \pm      1.59$ \\
4FGL J1028.3+3108  & 14.160 & $44.4 \pm  3.0$ & $ 2.23 \pm  0.15$ & 4.995 & $ 0.85 \pm  1.04$ & $  0.40 \pm 0.54$ & $+1.03 \pm 2.92$ & $ +1.20 \pm  3.07$ & $   0.62 \pm      1.60$ & $   0.49 \pm      1.27$ \\
4FGL J1031.3+5053  & 16.775 & $45.8 \pm  0.4$ & $ 1.74 \pm  0.03$ & 4.996 & $ 3.45 \pm  0.63$ & $  4.95 \pm 0.34$ & $-0.41 \pm 0.19$ & $ +2.15 \pm  0.23$ & $   1.51 \pm      0.38$ & $   1.90 \pm      0.48$ \\
4FGL J1033.5+4221  & 15.450 & $44.6 \pm  0.9$ & $ 1.76 \pm  0.22$ & 4.996 & $ 2.62 \pm  0.79$ & $  0.61 \pm 0.48$ & $+0.70 \pm 1.88$ & $ +1.06 \pm  2.10$ & $   0.29 \pm      0.31$ & $   0.26 \pm      0.28$ \\
4FGL J1049.7+5011  & 16.640 & $45.1 \pm  2.2$ & $ 2.25 \pm  0.11$ & 4.996 & $ 1.03 \pm  0.53$ & $  0.23 \pm 0.19$ & $+0.41 \pm 2.58$ & $ +1.83 \pm  2.69$ & $   0.29 \pm      0.38$ & $   0.26 \pm      0.34$ \\
4FGL J1051.9+0103  & 14.600 & $44.4 \pm  2.8$ & $ 1.93 \pm  0.11$ & 4.996 & $ 0.47 \pm  0.66$ & $  0.44 \pm 0.33$ & $+0.82 \pm 1.92$ & $ +1.11 \pm  2.03$ & $   1.20 \pm      2.59$ & $   0.99 \pm      2.13$ \\
4FGL J1053.7+4930  & 14.105 & $44.4 \pm  0.9$ & $ 1.92 \pm  0.06$ & 4.995 & $ 3.08 \pm  1.64$ & $  2.28 \pm 1.25$ & $+0.65 \pm 1.16$ & $ +1.27 \pm  1.22$ & $   0.88 \pm      0.95$ & $   0.81 \pm      0.88$ \\
4FGL J1058.6+5627  & 14.750 & $45.1 \pm  0.1$ & $ 1.94 \pm  0.02$ & 4.996 & $ 8.93 \pm  0.95$ & $ 13.66 \pm 0.16$ & $-0.61 \pm 0.35$ & $ +2.55 \pm  0.37$ & $   0.84 \pm      0.20$ & $   1.04 \pm      0.25$ \\
4FGL J1104.0+0020  & 15.300 & $44.8 \pm  1.1$ & $ 2.53 \pm  0.19$ & 4.996 & $ 1.46 \pm  0.65$ & $  0.73 \pm 0.28$ & $+1.66 \pm 0.75$ & $ +0.88 \pm  0.94$ & $   0.74 \pm      0.62$ & $   0.49 \pm      0.40$ \\
4FGL J1104.4+3812  & 16.220 & $44.6 \pm  0.4$ & $ 1.78 \pm  0.00$ & 4.996 & $116.40 \pm 60.06$ & $235.40 \pm 71.97$ & $-0.52 \pm 0.76$ & $ +2.30 \pm  0.76$ & $   2.40 \pm      1.98$ & $   2.74 \pm      2.25$ \\
4FGL J1105.8+3944  & 12.410 & $43.9 \pm  0.4$ & $ 2.22 \pm  0.19$ & 4.996 & $ 4.03 \pm  0.61$ & $  0.09 \pm 0.49$ & $+1.54 \pm 7.13$ & $ +0.68 \pm  7.32$ & $   0.03 \pm      0.15$ & $   0.02 \pm      0.12$ \\
4FGL J1107.8+1501  & 14.560 & $44.5 \pm  0.5$ & $ 1.95 \pm  0.06$ & 4.995 & $ 0.12 \pm  0.44$ & $  2.77 \pm 0.25$ & $+1.22 \pm 0.17$ & $ +0.73 \pm  0.22$ & $  30.58 \pm    116.90$ & $  24.10 \pm     92.11$ \\
4FGL J1109.6+3735  & 14.600 & $45.2 \pm  3.8$ & $ 1.85 \pm  0.13$ & 4.996 & $ 0.55 \pm  1.03$ & $  0.81 \pm 0.37$ & $+0.65 \pm 1.31$ & $ +1.21 \pm  1.44$ & $   1.98 \pm      4.56$ & $   1.61 \pm      3.71$ \\
4FGL J1112.4+1751  & 16.580 & $45.5 \pm  2.2$ & $ 2.30 \pm  0.17$ & 4.996 & $ 0.65 \pm  1.05$ & $  1.85 \pm 0.32$ & $+0.90 \pm 0.54$ & $ +1.41 \pm  0.71$ & $   4.18 \pm      7.52$ & $   3.04 \pm      5.47$ \\
4FGL J1117.0+2013  & 16.200 & $44.6 \pm  0.9$ & $ 1.96 \pm  0.04$ & 4.995 & $ 2.82 \pm  2.51$ & $  5.37 \pm 1.92$ & $+0.67 \pm 0.74$ & $ +1.28 \pm  0.78$ & $   2.27 \pm      2.83$ & $   2.09 \pm      2.61$ \\
4FGL J1117.2+0008  & 16.460 & $45.4 \pm  2.3$ & $ 1.88 \pm  0.15$ & 4.996 & $ 0.62 \pm  0.58$ & $  0.78 \pm 0.17$ & $+0.71 \pm 0.70$ & $ +1.17 \pm  0.85$ & $   1.77 \pm      2.04$ & $   1.37 \pm      1.57$ \\
4FGL J1120.8+4212  & 16.320 & $46.6 \pm  0.1$ & $ 1.62 \pm  0.03$ & 4.995 & undetected        & $  9.79 \pm 0.15$ & $+0.62 \pm 0.05$ & $ +1.00 \pm  0.08$ & --                      & --                      \\
4FGL J1131.4+5809  & 14.150 & $45.3 \pm  2.8$ & $ 2.07 \pm  0.09$ & 4.996 & $ 1.72 \pm  1.36$ & $  0.89 \pm 0.76$ & $-0.57 \pm 2.45$ & $ +2.64 \pm  2.54$ & $   0.52 \pm      0.86$ & $   0.71 \pm      1.16$ \\
4FGL J1136.4+6736  & 17.480 & $44.4 \pm  0.3$ & $ 1.75 \pm  0.05$ & 4.996 & $ 3.19 \pm  0.61$ & $  2.80 \pm 0.44$ & $+1.28 \pm 0.26$ & $ +0.46 \pm  0.31$ & $   1.06 \pm      0.37$ & $   0.89 \pm      0.31$ \\
4FGL J1136.8+2550  & 13.520 & $44.4 \pm  0.5$ & $ 1.94 \pm  0.11$ & 4.995 & $ 2.41 \pm  0.79$ & $  2.58 \pm 0.53$ & $+1.32 \pm 0.32$ & $ +0.62 \pm  0.43$ & $   1.33 \pm      0.71$ & $   1.08 \pm      0.58$ \\
4FGL J1140.5+1528  & 16.340 & $45.0 \pm  0.6$ & $ 1.76 \pm  0.15$ & 4.996 & $ 2.73 \pm  0.74$ & $  2.03 \pm 0.43$ & $+0.23 \pm 0.58$ & $ +1.53 \pm  0.73$ & $   0.89 \pm      0.43$ & $   0.88 \pm      0.42$ \\
4FGL J1145.5--0340 & 17.360 & $44.3 \pm  1.2$ & $ 1.93 \pm  0.21$ & 4.996 & $ 2.15 \pm  1.07$ & $  0.67 \pm 0.78$ & $+0.38 \pm 2.82$ & $ +1.55 \pm  3.02$ & $   0.37 \pm      0.61$ & $   0.36 \pm      0.59$ \\
4FGL J1149.4+2441  & 15.560 & $45.5 \pm  1.0$ & $ 1.96 \pm  0.14$ & 4.995 & $ 1.63 \pm  0.53$ & $  1.04 \pm 0.20$ & $+0.01 \pm 0.63$ & $ +1.95 \pm  0.78$ & $   0.74 \pm      0.38$ & $   0.78 \pm      0.40$ \\
\hline
\end{tabular}
\end{center}
\end{table}
\clearpage
\begin{table}
\contcaption{}
\begin{center}
\begin{tabular}{lcccccccccc}
\hline
\hline
Object name        & $\log \nu_{Peak}^{Syn}$ & $\log \nu L_\nu$                & $\alpha_\gamma$ & $\nu_\mathrm{norm}$ & Host norm.         & PL norm. & $\alpha_{opt}$ & $\alpha_\gamma - \alpha_{opt}$ & $f_{ratio}$ & $f_{ratio}^{int}$ \\
                   & Hz                      & $\mathrm{erg\, s^{-1}}$ &               & $10^{14}\,$Hz      & $10^{-4}\,$Jy       & $10^{-4}\,$Jy    &               &                             &           &                \\
\hline
4FGL J1152.1+2837  & 14.440 & $45.5 \pm  3.0$ & $ 1.88 \pm  0.18$ & 4.995 & $ 1.34 \pm  1.12$ & $  0.69 \pm 0.30$ & $+1.21 \pm 1.34$ & $ +0.68 \pm  1.52$ & $   0.82 \pm      1.05$ & $   0.52 \pm      0.66$ \\
4FGL J1153.7+3822  & 13.775 & $45.3 \pm  2.2$ & $ 1.93 \pm  0.17$ & 4.996 & $ 0.98 \pm  0.80$ & $  0.87 \pm 0.24$ & $+1.07 \pm 0.84$ & $ +0.86 \pm  1.01$ & $   1.34 \pm      1.47$ & $   0.92 \pm      1.01$ \\
4FGL J1154.0--0010 & 16.415 & $44.7 \pm  1.7$ & $ 1.81 \pm  0.08$ & 4.996 & $ 1.22 \pm  0.88$ & $  0.78 \pm 0.47$ & $+0.62 \pm 1.56$ & $ +1.18 \pm  1.64$ & $   0.81 \pm      1.07$ & $   0.71 \pm      0.94$ \\
4FGL J1202.4+4442  & 14.140 & $45.1 \pm  1.3$ & $ 2.60 \pm  0.18$ & 4.995 & $ 1.17 \pm  0.99$ & $  1.96 \pm 0.41$ & $+1.34 \pm 0.49$ & $ +1.26 \pm  0.67$ & $   2.44 \pm      2.58$ & $   1.69 \pm      1.79$ \\
4FGL J1203.1+6031  & 13.612 & $44.1 \pm  0.2$ & $ 2.12 \pm  0.04$ & 4.995 & $ 5.68 \pm  1.56$ & $ 15.65 \pm 1.43$ & $+1.18 \pm 0.14$ & $ +0.95 \pm  0.18$ & $   3.09 \pm      1.13$ & $   2.83 \pm      1.04$ \\
4FGL J1203.4--3925 & 15.800 & $44.5 \pm  1.4$ & $ 1.81 \pm  0.09$ & 4.996 & $ 1.04 \pm  0.75$ & $  0.91 \pm 0.43$ & $+0.61 \pm 1.32$ & $ +1.20 \pm  1.40$ & $   1.08 \pm      1.28$ & $   0.96 \pm      1.14$ \\
4FGL J1208.4+6121  & 14.076 & $45.1 \pm  0.6$ & $ 2.07 \pm  0.14$ & 4.995 & $ 2.87 \pm  0.57$ & $  1.03 \pm 0.32$ & $-0.12 \pm 0.92$ & $ +2.19 \pm  1.05$ & $   0.41 \pm      0.21$ & $   0.45 \pm      0.23$ \\
4FGL J1212.0+2242  & 16.635 & $45.4 \pm  2.8$ & $ 2.02 \pm  0.13$ & 4.996 & $ 0.47 \pm  0.87$ & $  1.21 \pm 0.24$ & $+0.88 \pm 0.65$ & $ +1.13 \pm  0.78$ & $   3.78 \pm      7.72$ & $   2.71 \pm      5.54$ \\
4FGL J1215.1+0731  & 13.908 & $44.4 \pm  0.3$ & $ 1.72 \pm  0.10$ & 4.996 & $ 4.04 \pm  0.50$ & $  1.84 \pm 0.37$ & $+1.22 \pm 0.39$ & $ +0.51 \pm  0.49$ & $   0.55 \pm      0.18$ & $   0.47 \pm      0.15$ \\
4FGL J1216.1+0930  & 15.350 & $44.0 \pm  0.2$ & $ 2.06 \pm  0.09$ & 4.996 & $ 4.24 \pm  0.41$ & $  2.49 \pm 0.33$ & $+2.10 \pm 0.09$ & $ -0.04 \pm  0.18$ & $   0.68 \pm      0.16$ & $   0.56 \pm      0.13$ \\
4FGL J1218.0--0028 & 13.640 & $45.6 \pm  2.7$ & $ 2.32 \pm  0.08$ & 4.995 & $ 1.28 \pm  1.53$ & $  1.63 \pm 0.40$ & $+1.79 \pm 0.63$ & $ +0.53 \pm  0.71$ & $   2.35 \pm      3.38$ & $   1.23 \pm      1.77$ \\
4FGL J1219.7--0313 & 14.360 & $45.2 \pm  1.6$ & $ 1.94 \pm  0.07$ & 4.996 & $ 1.03 \pm  1.46$ & $  2.68 \pm 0.63$ & $+1.10 \pm 0.58$ & $ +0.84 \pm  0.66$ & $   3.63 \pm      5.96$ & $   2.69 \pm      4.42$ \\
4FGL J1221.3+3010  & 16.270 & $45.1 \pm  0.4$ & $ 1.71 \pm  0.02$ & 4.995 & $ 2.36 \pm  1.71$ & $ 11.22 \pm 1.12$ & $+0.71 \pm 0.23$ & $ +1.00 \pm  0.25$ & $   5.82 \pm      4.79$ & $   5.19 \pm      4.27$ \\
4FGL J1223.8+4649  & 15.425 & $44.8 \pm  0.7$ & $ 2.22 \pm  0.19$ & 4.995 & $ 1.36 \pm  0.68$ & $  1.33 \pm 0.08$ & $+1.81 \pm 0.72$ & $ +0.41 \pm  0.92$ & $   0.80 \pm      0.74$ & $   0.52 \pm      0.48$ \\
4FGL J1224.4+2436  & 15.550 & $45.0 \pm  0.5$ & $ 1.93 \pm  0.04$ & 4.995 & $ 0.69 \pm  1.03$ & $  6.12 \pm 0.60$ & $+0.79 \pm 0.22$ & $ +1.14 \pm  0.26$ & $  11.17 \pm     17.78$ & $   9.58 \pm     15.25$ \\
4FGL J1230.9+3711  & 14.884 & $44.7 \pm  1.3$ & $ 2.51 \pm  0.24$ & 4.996 & $ 2.65 \pm  1.17$ & $  0.58 \pm 0.69$ & $+1.05 \pm 2.97$ & $ +1.45 \pm  3.21$ & $   0.28 \pm      0.46$ & $   0.23 \pm      0.37$ \\
4FGL J1231.5+1421  & 14.765 & $45.1 \pm  0.7$ & $ 1.88 \pm  0.10$ & 4.995 & $ 1.98 \pm  0.91$ & $  2.78 \pm 0.45$ & $+1.07 \pm 0.38$ & $ +0.81 \pm  0.48$ & $   1.88 \pm      1.17$ & $   1.46 \pm      0.91$ \\
4FGL J1231.6+6415  & 14.040 & $44.6 \pm  0.3$ & $ 1.91 \pm  0.08$ & 4.996 & $ 4.53 \pm  0.63$ & $  1.67 \pm 0.50$ & $-0.08 \pm 0.78$ & $ +1.99 \pm  0.86$ & $   0.42 \pm      0.19$ & $   0.46 \pm      0.20$ \\
4FGL J1233.6+5027  & 14.576 & $44.6 \pm  1.2$ & $ 2.46 \pm  0.11$ & 4.995 & $ 1.89 \pm  0.99$ & $  0.92 \pm 0.56$ & $+1.31 \pm 1.19$ & $ +1.15 \pm  1.30$ & $   0.64 \pm      0.72$ & $   0.49 \pm      0.56$ \\
4FGL J1236.3+3858  & 14.090 & $45.2 \pm  4.3$ & $ 1.99 \pm  0.12$ & 4.995 & $ 0.86 \pm  1.52$ & $  0.85 \pm 0.49$ & $+1.08 \pm 1.68$ & $ +0.91 \pm  1.80$ & $   1.48 \pm      3.45$ & $   1.03 \pm      2.40$ \\
4FGL J1246.3+0112  & 13.472 & $45.3 \pm  2.8$ & $ 2.10 \pm  0.17$ & 4.995 & $ 1.08 \pm  1.14$ & $  0.87 \pm 0.34$ & $+1.54 \pm 1.03$ & $ +0.56 \pm  1.20$ & $   1.34 \pm      1.93$ & $   0.79 \pm      1.14$ \\
4FGL J1248.7+5127  & 14.325 & $45.3 \pm  1.8$ & $ 2.07 \pm  0.11$ & 4.996 & $ 1.58 \pm  1.23$ & $  1.56 \pm 0.46$ & $+0.98 \pm 0.76$ & $ +1.09 \pm  0.87$ & $   1.41 \pm      1.51$ & $   1.04 \pm      1.12$ \\
4FGL J1251.2+1039  & 15.315 & $44.8 \pm  1.3$ & $ 2.13 \pm  0.13$ & 4.996 & $ 0.70 \pm  1.06$ & $  2.27 \pm 0.54$ & $+1.03 \pm 0.55$ & $ +1.10 \pm  0.69$ & $   4.28 \pm      7.47$ & $   3.38 \pm      5.90$ \\
4FGL J1253.8+0327  & 15.920 & $44.0 \pm  0.2$ & $ 1.99 \pm  0.05$ & 4.996 & $ 9.82 \pm  1.29$ & $  5.12 \pm 1.16$ & $+1.38 \pm 0.29$ & $ +0.61 \pm  0.34$ & $   0.58 \pm      0.21$ & $   0.52 \pm      0.19$ \\
4FGL J1257.2+3646  & 14.900 & $46.2 \pm  1.0$ & $ 2.03 \pm  0.06$ & 4.996 & $ 2.00 \pm  0.81$ & $  3.63 \pm 0.21$ & $+0.58 \pm 0.19$ & $ +1.44 \pm  0.24$ & $   2.56 \pm      1.18$ & $   2.02 \pm      0.94$ \\
4FGL J1257.6+2413  & 18.680 & $44.3 \pm  0.9$ & $ 1.65 \pm  0.22$ & 4.996 & $ 2.85 \pm  1.21$ & $  1.24 \pm 0.95$ & $+0.41 \pm 1.78$ & $ +1.24 \pm  2.01$ & $   0.51 \pm      0.61$ & $   0.50 \pm      0.59$ \\
4FGL J1304.0+3704  & 13.543 & $46.6 \pm 12.3$ & $ 2.44 \pm  0.14$ & 4.995 & $ 0.45 \pm  1.14$ & $  0.93 \pm 0.08$ & $+1.35 \pm 0.34$ & $ +1.08 \pm  0.48$ & $   5.22 \pm     13.67$ & $   2.09 \pm      5.48$ \\
4FGL J1319.5+1404  & 14.982 & $46.3 \pm  1.1$ & $ 2.00 \pm  0.12$ & 4.995 & $ 1.51 \pm  0.69$ & $  2.99 \pm 0.14$ & $+0.68 \pm 0.17$ & $ +1.32 \pm  0.29$ & $   2.92 \pm      1.46$ & $   2.16 \pm      1.08$ \\
4FGL J1322.9+0437  & 16.030 & $44.7 \pm  1.3$ & $ 1.92 \pm  0.13$ & 4.996 & $ 2.46 \pm  1.20$ & $  0.82 \pm 0.71$ & $+0.58 \pm 2.20$ & $ +1.34 \pm  2.33$ & $   0.41 \pm      0.56$ & $   0.37 \pm      0.50$ \\
4FGL J1326.1+1232  & 16.690 & $44.6 \pm  0.7$ & $ 1.94 \pm  0.19$ & 4.996 & $ 1.78 \pm  0.51$ & $  1.11 \pm 0.38$ & $-0.01 \pm 0.88$ & $ +1.95 \pm  1.06$ & $   0.72 \pm      0.45$ & $   0.76 \pm      0.48$ \\
4FGL J1336.2+2320  & 15.020 & $44.8 \pm  1.5$ & $ 2.00 \pm  0.14$ & 4.995 & $ 1.23 \pm  0.85$ & $  0.93 \pm 0.42$ & $+0.75 \pm 1.18$ & $ +1.24 \pm  1.32$ & $   0.97 \pm      1.11$ & $   0.82 \pm      0.93$ \\
4FGL J1340.5+4409  & 16.085 & $45.7 \pm  4.2$ & $ 1.66 \pm  0.13$ & 4.996 & $ 0.41 \pm  0.94$ & $  1.09 \pm 0.18$ & $+1.37 \pm 0.51$ & $ +0.29 \pm  0.65$ & $   4.87 \pm     11.86$ & $   2.64 \pm      6.42$ \\
4FGL J1340.8--0409 & 15.320 & $45.2 \pm  0.4$ & $ 2.01 \pm  0.09$ & 4.996 & $ 2.53 \pm  1.06$ & $  6.07 \pm 0.62$ & $+0.74 \pm 0.25$ & $ +1.27 \pm  0.34$ & $   3.02 \pm      1.58$ & $   2.61 \pm      1.37$ \\
4FGL J1341.6+5515  & 14.390 & $44.6 \pm  1.1$ & $ 2.09 \pm  0.19$ & 4.996 & $ 1.78 \pm  0.96$ & $  1.11 \pm 0.57$ & $+1.01 \pm 1.12$ & $ +1.08 \pm  1.31$ & $   0.79 \pm      0.83$ & $   0.65 \pm      0.69$ \\
4FGL J1353.3+1434  & 13.490 & $46.5 \pm  4.6$ & $ 2.25 \pm  0.07$ & 4.996 & $ 1.00 \pm  0.84$ & $  1.11 \pm 0.08$ & $+1.22 \pm 0.30$ & $ +1.03 \pm  0.36$ & $   2.38 \pm      2.19$ & $   1.13 \pm      1.04$ \\
4FGL J1353.4+5600  & 14.670 & $45.4 \pm  1.8$ & $ 2.18 \pm  0.17$ & 4.996 & $ 1.07 \pm  0.68$ & $  0.94 \pm 0.24$ & $+0.54 \pm 0.77$ & $ +1.64 \pm  0.93$ & $   1.16 \pm      1.05$ & $   0.99 \pm      0.89$ \\
4FGL J1354.7+0623  & 14.400 & $44.7 \pm  2.2$ & $ 2.07 \pm  0.18$ & 4.995 & $ 0.71 \pm  0.81$ & $  0.72 \pm 0.39$ & $+0.74 \pm 1.44$ & $ +1.33 \pm  1.62$ & $   1.31 \pm      2.19$ & $   1.10 \pm      1.84$ \\
4FGL J1402.6+1600  & 14.450 & $44.8 \pm  1.3$ & $ 1.96 \pm  0.17$ & 4.995 & $ 0.95 \pm  1.02$ & $  1.89 \pm 0.52$ & $+1.16 \pm 0.61$ & $ +0.80 \pm  0.78$ & $   2.67 \pm      3.59$ & $   2.05 \pm      2.75$ \\
4FGL J1403.4+4319  & 14.215 & $45.4 \pm  7.1$ & $ 1.56 \pm  0.20$ & 4.995 & $ 0.71 \pm  1.18$ & $  0.31 \pm 0.27$ & $+1.24 \pm 2.65$ & $ +0.32 \pm  2.85$ & $   0.73 \pm      1.86$ & $   0.44 \pm      1.12$ \\
4FGL J1404.8+6554  & 12.930 & $45.3 \pm  4.3$ & $ 1.93 \pm  0.06$ & 4.995 & $ 1.54 \pm  2.26$ & $  1.10 \pm 1.12$ & $-0.15 \pm 2.95$ & $ +2.08 \pm  3.01$ & $   0.80 \pm      1.97$ & $   0.90 \pm      2.23$ \\
4FGL J1417.9+2543  & 17.060 & $45.1 \pm  0.5$ & $ 1.45 \pm  0.09$ & 4.996 & $ 2.91 \pm  0.86$ & $  3.33 \pm 0.51$ & $+0.40 \pm 0.40$ & $ +1.05 \pm  0.49$ & $   1.39 \pm      0.62$ & $   1.31 \pm      0.59$ \\
4FGL J1419.3+0444  & 14.860 & $44.0 \pm  0.1$ & $ 1.91 \pm  0.12$ & 4.995 & undetected        & $  1.85 \pm 0.06$ & $+0.92 \pm 0.13$ & $ +0.99 \pm  0.26$ & --                      & --                      \\
4FGL J1419.8+5423  & 13.680 & $45.2 \pm  0.2$ & $ 2.36 \pm  0.03$ & 4.996 & $ 2.03 \pm  1.79$ & $ 24.06 \pm 1.22$ & $+1.30 \pm 0.09$ & $ +1.07 \pm  0.12$ & $  14.68 \pm     13.68$ & $  12.01 \pm     11.19$ \\
4FGL J1427.0+2348  & 15.293 & $47.5 \pm  0.3$ & $ 1.82 \pm  0.01$ & 4.996 & $ 2.38 \pm  2.53$ & $ 55.81 \pm 0.35$ & $+1.10 \pm 0.03$ & $ +0.72 \pm  0.04$ & $  41.22 \pm     44.04$ & $  24.29 \pm     25.95$ \\
4FGL J1428.5+4240  & 18.010 & $44.5 \pm  0.2$ & $ 1.63 \pm  0.04$ & 4.996 & $ 6.14 \pm  0.66$ & $  3.24 \pm 0.70$ & $-0.65 \pm 0.54$ & $ +2.29 \pm  0.59$ & $   0.60 \pm      0.19$ & $   0.74 \pm      0.24$ \\
4FGL J1435.5+2021  & 13.715 & $46.5 \pm  5.7$ & $ 2.22 \pm  0.10$ & 4.996 & $ 0.09 \pm  1.64$ & $  2.80 \pm 0.19$ & $+1.18 \pm 0.26$ & $ +1.03 \pm  0.37$ & $  65.77 \pm   1266.00$ & $  33.54 \pm    645.40$ \\
\hline
\end{tabular}
\end{center}
\end{table}
\clearpage
\begin{table}
\contcaption{}
\begin{center}
\begin{tabular}{lcccccccccc}
\hline
\hline
Object name        & $\log \nu_{Peak}^{Syn}$ & $\log \nu L_\nu$                & $\alpha_\gamma$ & $\nu_\mathrm{norm}$ & Host norm.         & PL norm. & $\alpha_{opt}$ & $\alpha_\gamma - \alpha_{opt}$ & $f_{ratio}$ & $f_{ratio}^{int}$ \\
                   & Hz                      & $\mathrm{erg\, s^{-1}}$ &               & $10^{14}\,$Hz      & $10^{-4}\,$Jy       & $10^{-4}\,$Jy    &               &                             &           &                \\
\hline
4FGL J1439.9--3953 & 15.650 & $44.8 \pm  1.0$ & $ 2.09 \pm  0.11$ & 4.996 & $ 0.81 \pm  0.36$ & $  0.66 \pm 0.16$ & $+0.91 \pm 0.68$ & $ +1.18 \pm  0.79$ & $   1.10 \pm      0.75$ & $   0.87 \pm      0.59$ \\
4FGL J1442.7+1200  & 16.905 & $44.6 \pm  0.6$ & $ 1.80 \pm  0.06$ & 4.996 & $ 2.78 \pm  1.10$ & $  3.10 \pm 0.78$ & $+0.68 \pm 0.56$ & $ +1.12 \pm  0.63$ & $   1.34 \pm      0.87$ & $   1.22 \pm      0.79$ \\
4FGL J1503.5+4759  & 13.115 & $45.4 \pm  1.4$ & $ 2.22 \pm  0.06$ & 4.996 & $ 1.31 \pm  1.15$ & $  2.28 \pm 0.43$ & $+1.06 \pm 0.47$ & $ +1.16 \pm  0.54$ & $   2.50 \pm      2.66$ & $   1.81 \pm      1.93$ \\
4FGL J1506.4+4331  & 14.280 & $45.4 \pm  5.1$ & $ 2.30 \pm  0.19$ & 4.996 & $ 0.21 \pm  1.08$ & $  0.98 \pm 0.26$ & $+1.38 \pm 0.73$ & $ +0.92 \pm  0.92$ & $   8.02 \pm     42.68$ & $   4.63 \pm     24.64$ \\
4FGL J1507.2+1721  & 14.600 & $45.7 \pm  5.3$ & $ 1.81 \pm  0.08$ & 4.996 & $ 0.29 \pm  0.98$ & $  1.01 \pm 0.20$ & $+0.84 \pm 0.68$ & $ +0.97 \pm  0.76$ & $   5.42 \pm     19.20$ & $   3.71 \pm     13.15$ \\
4FGL J1508.8+2708  & 16.195 & $45.0 \pm  1.2$ & $ 2.05 \pm  0.12$ & 4.996 & $ 1.76 \pm  1.00$ & $  1.45 \pm 0.51$ & $+0.61 \pm 0.93$ & $ +1.43 \pm  1.05$ & $   1.04 \pm      0.96$ & $   0.91 \pm      0.84$ \\
4FGL J1509.7+5556  & 14.150 & $48.1 \pm 14.7$ & $ 1.88 \pm  0.07$ & 4.996 & $ 3.51 \pm  2.97$ & $  3.02 \pm 0.08$ & $+0.43 \pm 0.09$ & $ +1.45 \pm  0.16$ & $   1.39 \pm      1.21$ & $   0.98 \pm      0.86$ \\
4FGL J1518.6+4044  & 15.700 & $43.7 \pm  0.4$ & $ 1.88 \pm  0.09$ & 4.996 & $ 7.12 \pm  1.26$ & $  0.65 \pm 1.15$ & $+1.33 \pm 2.38$ & $ +0.54 \pm  2.46$ & $   0.10 \pm      0.20$ & $   0.09 \pm      0.18$ \\
4FGL J1522.6--2730 & 12.725 & $48.0 \pm  8.7$ & $ 2.20 \pm  0.03$ & 4.996 & $ 3.64 \pm  2.26$ & $  3.60 \pm 0.03$ & $+1.52 \pm 0.03$ & $ +0.68 \pm  0.05$ & $   3.55 \pm      2.22$ & $   0.98 \pm      0.62$ \\
4FGL J1523.2+0533  & 12.575 & $44.4 \pm  0.8$ & $ 2.03 \pm  0.16$ & 4.996 & $ 3.07 \pm  0.83$ & $  0.44 \pm 0.56$ & $+0.69 \pm 2.94$ & $ +1.34 \pm  3.11$ & $   0.17 \pm      0.27$ & $   0.16 \pm      0.24$ \\
4FGL J1532.0+3016  & 14.270 & $43.7 \pm  0.3$ & $ 1.92 \pm  0.07$ & 4.996 & $ 5.53 \pm  1.04$ & $  2.45 \pm 0.94$ & $+1.32 \pm 0.52$ & $ +0.60 \pm  0.59$ & $   0.50 \pm      0.28$ & $   0.45 \pm      0.26$ \\
4FGL J1533.2+1855  & 14.520 & $45.1 \pm  0.9$ & $ 1.82 \pm  0.10$ & 4.995 & $ 1.64 \pm  0.57$ & $  1.20 \pm 0.28$ & $+0.29 \pm 0.67$ & $ +1.53 \pm  0.77$ & $   0.89 \pm      0.52$ & $   0.85 \pm      0.49$ \\
4FGL J1534.8+3716  & 14.200 & $44.3 \pm  0.5$ & $ 2.06 \pm  0.11$ & 4.996 & $ 1.45 \pm  0.79$ & $  3.00 \pm 0.57$ & $+1.15 \pm 0.34$ & $ +0.90 \pm  0.45$ & $   2.52 \pm      1.86$ & $   2.13 \pm      1.57$ \\
4FGL J1535.4+3919  & 14.160 & $45.1 \pm  0.8$ & $ 2.20 \pm  0.10$ & 4.996 & $ 2.03 \pm  1.17$ & $  3.27 \pm 0.57$ & $+1.16 \pm 0.36$ & $ +1.04 \pm  0.47$ & $   2.18 \pm      1.64$ & $   1.65 \pm      1.24$ \\
4FGL J1540.7+1449  & 13.325 & $46.0 \pm  5.3$ & $ 2.15 \pm  0.10$ & 4.996 & $ 0.63 \pm  1.46$ & $  1.49 \pm 0.22$ & $+1.48 \pm 0.50$ & $ +0.66 \pm  0.60$ & $   4.85 \pm     12.01$ & $   2.36 \pm      5.84$ \\
4FGL J1541.7+1413  & 12.928 & $44.8 \pm  0.9$ & $ 1.96 \pm  0.13$ & 4.996 & $ 1.23 \pm  1.03$ & $  2.60 \pm 0.56$ & $+1.19 \pm 0.45$ & $ +0.77 \pm  0.59$ & $   2.79 \pm      2.92$ & $   2.16 \pm      2.27$ \\
4FGL J1548.3+1456  & 14.250 & $44.6 \pm  0.6$ & $ 2.05 \pm  0.03$ & 4.996 & $ 1.30 \pm  0.47$ & $  0.81 \pm 0.05$ & $+1.50 \pm 3.14$ & $ +0.55 \pm  3.17$ & $   0.18 \pm      0.33$ & $   0.13 \pm      0.24$ \\
4FGL J1606.2+1346  & 15.800 & $45.2 \pm  0.9$ & $ 1.76 \pm  0.18$ & 4.996 & $ 2.17 \pm  0.91$ & $  2.05 \pm 0.42$ & $+0.73 \pm 0.56$ & $ +1.02 \pm  0.74$ & $   1.23 \pm      0.77$ & $   1.03 \pm      0.64$ \\
4FGL J1616.7+4107  & 13.400 & $44.8 \pm  0.8$ & $ 2.25 \pm  0.12$ & 4.996 & $ 1.12 \pm  0.56$ & $  1.36 \pm 0.25$ & $+1.50 \pm 0.33$ & $ +0.75 \pm  0.45$ & $   1.76 \pm      1.20$ & $   1.21 \pm      0.82$ \\
4FGL J1626.3+3514  & 15.740 & $45.7 \pm  2.7$ & $ 1.75 \pm  0.11$ & 4.996 & $ 1.12 \pm  0.80$ & $  0.81 \pm 0.26$ & $+0.15 \pm 0.99$ & $ +1.60 \pm  1.10$ & $   0.88 \pm      0.90$ & $   0.87 \pm      0.89$ \\
4FGL J1626.6--7639 & 14.780 & $44.5 \pm  0.3$ & $ 2.04 \pm  0.08$ & 4.996 & $ 4.32 \pm  2.49$ & $  9.76 \pm 0.18$ & $+0.37 \pm 0.81$ & $ +1.67 \pm  0.90$ & $   1.64 \pm      1.53$ & $   1.63 \pm      1.52$ \\
4FGL J1637.1+1316  & 14.600 & $46.1 \pm  0.9$ & $ 2.26 \pm  0.16$ & 4.996 & $ 0.05 \pm  0.17$ & $  1.51 \pm 0.02$ & $+0.85 \pm 0.06$ & $ +1.41 \pm  0.22$ & $  46.61 \pm    150.30$ & $  30.45 \pm     98.17$ \\
4FGL J1637.2+4327  & 15.040 & $45.2 \pm  2.0$ & $ 1.80 \pm  0.20$ & 4.995 & $ 1.52 \pm  1.20$ & $  1.16 \pm 0.46$ & $+0.91 \pm 1.11$ & $ +0.90 \pm  1.31$ & $   1.07 \pm      1.26$ & $   0.81 \pm      0.96$ \\
4FGL J1637.6+4548  & 15.040 & $44.5 \pm  0.7$ & $ 1.92 \pm  0.20$ & 4.996 & $ 2.18 \pm  0.61$ & $  0.81 \pm 0.36$ & $+1.35 \pm 0.82$ & $ +0.57 \pm  1.03$ & $   0.48 \pm      0.35$ & $   0.37 \pm      0.27$ \\
4FGL J1642.4+2211  & 15.335 & $45.9 \pm  2.6$ & $ 2.02 \pm  0.19$ & 4.996 & $ 0.96 \pm  0.61$ & $  0.80 \pm 0.12$ & $+0.47 \pm 0.56$ & $ +1.54 \pm  0.75$ & $   1.14 \pm      0.90$ & $   0.94 \pm      0.74$ \\
4FGL J1643.0+3223  & 13.560 & $44.8 \pm  2.7$ & $ 2.47 \pm  0.13$ & 4.996 & $ 0.15 \pm  0.49$ & $  0.70 \pm 0.15$ & $+1.73 \pm 0.43$ & $ +0.73 \pm  0.57$ & $   7.95 \pm     27.66$ & $   4.51 \pm     15.68$ \\
4FGL J1647.5+2911  & 14.480 & $44.2 \pm  0.4$ & $ 2.27 \pm  0.13$ & 4.996 & $ 3.01 \pm  0.53$ & $  1.22 \pm 0.37$ & $+1.47 \pm 0.41$ & $ +0.80 \pm  0.53$ & $   0.49 \pm      0.24$ & $   0.40 \pm      0.19$ \\
4FGL J1652.7+4024  & 15.435 & $48.7 \pm 89.5$ & $ 1.93 \pm  0.12$ & 4.995 & $ 4.93 \pm 10.25$ & $  2.10 \pm 0.08$ & $+1.11 \pm 0.13$ & $ +0.82 \pm  0.25$ & $   1.39 \pm      2.95$ & $   0.44 \pm      0.93$ \\
4FGL J1653.8+3945  & 15.450 & $44.1 \pm  1.0$ & $ 1.76 \pm  0.01$ & 4.996 & $55.51 \pm 32.74$ & $ 28.45 \pm 36.36$ & $+0.45 \pm 2.48$ & $ +1.31 \pm  2.49$ & $   0.58 \pm      1.08$ & $   0.58 \pm      1.09$ \\
4FGL J1706.8+3004  & 14.100 & $47.5 \pm 84.4$ & $ 2.14 \pm  0.19$ & 4.996 & $ 0.47 \pm  3.06$ & $  0.86 \pm 0.05$ & $+1.53 \pm 0.23$ & $ +0.62 \pm  0.42$ & $   7.24 \pm     47.23$ & $   1.80 \pm     11.73$ \\
4FGL J1733.4+5428  & 14.400 & $45.0 \pm  3.9$ & $ 2.88 \pm  0.12$ & 4.995 & $ 0.72 \pm  0.50$ & $  0.06 \pm 0.22$ & $-0.69 \pm 11.56$ & $ +3.57 \pm 11.68$ & $   0.08 \pm      0.35$ & $   0.12 \pm      0.51$ \\
4FGL J1954.9--5640 & 15.900 & $45.2 \pm  0.3$ & $ 1.91 \pm  0.08$ & 4.996 & $ 3.24 \pm  0.99$ & $  6.79 \pm 0.58$ & $+0.81 \pm 0.22$ & $ +1.10 \pm  0.29$ & $   2.65 \pm      1.04$ & $   2.26 \pm      0.88$ \\
4FGL J2000.0+6508  & 15.956 & $44.1 \pm  1.0$ & $ 1.82 \pm  0.01$ & 4.995 & $ 4.92 \pm 17.49$ & $ 35.72 \pm 17.85$ & $+0.94 \pm 0.87$ & $ +0.88 \pm  0.88$ & $   8.06 \pm     32.67$ & $   7.68 \pm     31.12$ \\
4FGL J2054.8+0015  & 12.795 & $44.4 \pm  0.5$ & $ 1.82 \pm  0.12$ & 4.996 & $ 2.89 \pm  0.77$ & $  1.45 \pm 0.54$ & $+1.10 \pm 0.70$ & $ +0.72 \pm  0.82$ & $   0.61 \pm      0.39$ & $   0.52 \pm      0.33$ \\
4FGL J2055.4--0504 & 14.480 & $45.1 \pm  2.9$ & $ 2.26 \pm  0.15$ & 4.996 & $ 1.49 \pm  1.34$ & $  0.53 \pm 0.49$ & $+1.22 \pm 2.41$ & $ +1.05 \pm  2.56$ & $   0.53 \pm      0.96$ & $   0.36 \pm      0.66$ \\
4FGL J2115.9--0113 & 13.960 & $45.1 \pm  1.6$ & $ 2.11 \pm  0.09$ & 4.995 & $ 1.05 \pm  1.22$ & $  2.12 \pm 0.51$ & $+1.11 \pm 0.61$ & $ +1.00 \pm  0.71$ & $   2.83 \pm      3.95$ & $   2.08 \pm      2.91$ \\
4FGL J2131.5--0916 & 16.466 & $45.8 \pm  0.4$ & $ 1.99 \pm  0.06$ & 4.996 & $ 1.24 \pm  0.32$ & $  2.78 \pm 0.09$ & $+0.43 \pm 0.10$ & $ +1.56 \pm  0.16$ & $   2.92 \pm      0.85$ & $   2.56 \pm      0.74$ \\
4FGL J2153.1--0041 & 14.480 & $45.3 \pm  1.6$ & $ 1.79 \pm  0.15$ & 4.995 & $ 0.63 \pm  1.02$ & $  2.13 \pm 0.40$ & $+0.85 \pm 0.53$ & $ +0.94 \pm  0.68$ & $   4.67 \pm      8.44$ & $   3.63 \pm      6.57$ \\
4FGL J2158.8--3013 & 15.755 & $45.8 \pm  0.1$ & $ 1.85 \pm  0.01$ & 4.996 & $19.85 \pm  9.67$ & $173.40 \pm 8.18$ & $-0.18 \pm 0.12$ & $ +2.03 \pm  0.13$ & $  10.04 \pm      5.37$ & $  11.07 \pm      5.92$ \\
4FGL J2159.1--2840 & 15.400 & $44.9 \pm  0.4$ & $ 1.88 \pm  0.08$ & 4.996 & $ 0.77 \pm  0.37$ & $  1.98 \pm 0.02$ & $+0.73 \pm 0.32$ & $ +1.14 \pm  0.39$ & $   2.67 \pm      1.58$ & $   2.25 \pm      1.33$ \\
4FGL J2202.7+4216  & 13.587 & $44.9 \pm  0.1$ & $ 2.20 \pm  0.01$ & 4.996 & $10.58 \pm  4.65$ & $112.30 \pm 4.65$ & $+1.99 \pm 0.06$ & $ +0.21 \pm  0.07$ & $  11.76 \pm      5.66$ & $  10.12 \pm      4.87$ \\
4FGL J2206.8--0032 & 12.840 & $46.7 \pm 39.8$ & $ 2.25 \pm  0.05$ & 4.996 & $ 0.25 \pm  1.92$ & $  0.63 \pm 0.04$ & $+1.58 \pm 0.50$ & $ +0.67 \pm  0.55$ & $   7.84 \pm     61.90$ & $   2.48 \pm     19.55$ \\
4FGL J2211.0--0003 & 14.335 & $45.3 \pm  3.5$ & $ 1.95 \pm  0.13$ & 4.996 & $ 0.64 \pm  1.86$ & $  1.75 \pm 0.64$ & $+1.21 \pm 0.90$ & $ +0.74 \pm  1.03$ & $   4.13 \pm     13.55$ & $   2.80 \pm      9.19$ \\
4FGL J2220.5+2813  & 15.700 & $44.3 \pm  0.4$ & $ 1.99 \pm  0.11$ & 4.995 & $ 1.92 \pm  0.56$ & $  1.83 \pm 0.38$ & $+1.50 \pm 0.27$ & $ +0.48 \pm  0.39$ & $   1.19 \pm      0.59$ & $   0.95 \pm      0.47$ \\
\hline
\end{tabular}
\end{center}
\end{table}
\clearpage
\begin{table}
\contcaption{}
\begin{center}
\begin{tabular}{lcccccccccc}
\hline
\hline
Object name        & $\log \nu_{Peak}^{Syn}$ & $\log \nu L_\nu$                & $\alpha_\gamma$ & $\nu_\mathrm{norm}$ & Host norm.         & PL norm. & $\alpha_{opt}$ & $\alpha_\gamma - \alpha_{opt}$ & $f_{ratio}$ & $f_{ratio}^{int}$ \\
                   & Hz                      & $\mathrm{erg\, s^{-1}}$ &               & $10^{14}\,$Hz      & $10^{-4}\,$Jy       & $10^{-4}\,$Jy    &               &                             &           &                \\
\hline
4FGL J2225.5--1114 & 12.725 & $45.8 \pm 23.6$ & $ 2.27 \pm  0.14$ & 4.996 & $ 0.04 \pm  0.24$ & $  0.12 \pm 0.00$ & $+1.66 \pm 0.19$ & $ +0.61 \pm  0.33$ & $   9.24 \pm     54.30$ & $   2.87 \pm     16.89$ \\
4FGL J2228.6--1636 & 14.413 & $45.8 \pm  1.0$ & $ 2.06 \pm  0.10$ & 4.996 & $ 0.49 \pm  0.38$ & $  1.56 \pm 0.02$ & $+0.88 \pm 0.20$ & $ +1.18 \pm  0.30$ & $   4.68 \pm      3.93$ & $   3.23 \pm      2.71$ \\
4FGL J2232.8+1334  & 18.500 & $44.5 \pm  1.1$ & $ 1.78 \pm  0.19$ & 4.996 & $ 1.38 \pm  0.78$ & $  0.96 \pm 0.45$ & $+0.91 \pm 1.07$ & $ +0.87 \pm  1.26$ & $   0.88 \pm      0.91$ & $   0.74 \pm      0.76$ \\
4FGL J2244.9--0007 & 15.920 & $46.3 \pm  0.7$ & $ 2.31 \pm  0.18$ & 4.996 & $ 0.39 \pm  0.30$ & $  2.90 \pm 0.04$ & $+0.86 \pm 0.06$ & $ +1.46 \pm  0.24$ & $  12.19 \pm      9.62$ & $   7.98 \pm      6.30$ \\
4FGL J2245.9+1544  & 14.320 & $45.9 \pm  1.6$ & $ 1.96 \pm  0.08$ & 4.996 & $ 0.24 \pm  0.35$ & $  1.35 \pm 0.05$ & $+1.12 \pm 0.16$ & $ +0.85 \pm  0.23$ & $   9.83 \pm     14.69$ & $   5.76 \pm      8.60$ \\
4FGL J2247.4--0001 & 13.530 & $43.3 \pm  1.6$ & $ 2.06 \pm  0.08$ & 4.995 & $ 0.01 \pm  0.51$ & $  0.86 \pm 0.39$ & $+1.54 \pm 0.54$ & $ +0.52 \pm  0.62$ & $ 109.60 \pm   6131.00$ & $  91.81 \pm   5135.00$ \\
4FGL J2252.6+1245  & 14.435 & $45.7 \pm  4.1$ & $ 1.94 \pm  0.16$ & 4.996 & $ 0.80 \pm  1.27$ & $  1.10 \pm 0.29$ & $+1.08 \pm 0.89$ & $ +0.86 \pm  1.05$ & $   2.25 \pm      4.19$ & $   1.44 \pm      2.68$ \\
4FGL J2255.2+2411  & 13.928 & $48.3 \pm 16.8$ & $ 2.11 \pm  0.04$ & 4.995 & $ 4.88 \pm  6.41$ & $  7.26 \pm 0.13$ & $+1.40 \pm 0.08$ & $ +0.70 \pm  0.12$ & $   5.11 \pm      6.79$ & $   1.49 \pm      1.98$ \\
4FGL J2314.0+1445  & 17.075 & $44.6 \pm  0.2$ & $ 1.85 \pm  0.06$ & 4.996 & $ 4.23 \pm  0.30$ & $  1.25 \pm 0.30$ & $-0.90 \pm 0.62$ & $ +2.75 \pm  0.68$ & $   0.32 \pm      0.10$ & $   0.44 \pm      0.14$ \\
4FGL J2343.6+3438  & 17.600 & $45.2 \pm  1.9$ & $ 1.77 \pm  0.08$ & 4.996 & $ 1.09 \pm  0.84$ & $  1.10 \pm 0.35$ & $+0.38 \pm 0.92$ & $ +1.38 \pm  1.00$ & $   1.27 \pm      1.38$ & $   1.16 \pm      1.26$ \\
4FGL J2354.1+2720  & 13.670 & $43.8 \pm  0.7$ & $ 2.24 \pm  0.13$ & 4.996 & $ 0.37 \pm  0.44$ & $  0.68 \pm 0.04$ & $+0.27 \pm 1.99$ & $ +1.97 \pm  2.12$ & $   1.32 \pm      2.71$ & $   1.32 \pm      2.71$ \\
\hline
\end{tabular}
\end{center}
\end{table}
\end{footnotesize}
\end{landscape}

\appendix
\section{Synchrotron-Self Compton models}
\label{secModel}
The theory of synchrotron radiation can be fairly well described in the case of a distribution of relativistic particles that interact with a magnetic field with random inclinations \citep{Rybicki86}. In such case, we expect a total specific emission coefficient given by:
\begin{equation}
    j_\nu^{syn} = \dfrac{\sqrt{3} q^3 B}{m c^2} \int_{\gamma_{min}}^{\gamma_{max}} \de \gamma N(\gamma) x \int_x^\infty K_{5/3}(\xi) \de \xi, \label{eqnJsyn} 
\end{equation}
where $q$ and $m$ are the particle charge and mass, $B$ is the average magnetic field intensity, $K_{5/3}(\xi)$ is a modified Bessel function of the second type with order $5/3$, while $x = \nu / \nu_c$ expresses the ratio between the emission frequency $\nu$ and the critical synchrotron frequency for a particle with Lorentz factor $\gamma$:
\begin{equation}
    \nu_c = \dfrac{\gamma^2 q B}{2 \pi m c}. \label{eqnNuc}
\end{equation}
If the radiating particle density follows a power-law distribution of type $N(\gamma) = N_0 \gamma^{-p}$, integration over $\gamma$ leads to an emission coefficient $j_\nu^{syn} \propto \nu^{-(p-1)/2}$, while the synchrotron absorption coefficient takes the form:
\begin{equation}
\begin{array}{rcl}
    k_\nu^{syn} & = & \dfrac{\sqrt{3} N_0 q^3}{8 \pi m} \left( \dfrac{3 q}{2 \pi m^3 c^5} \right)^{p / 2} \left( \dfrac{2}{3} B \right)^{(p + 2) / 2} \times \\
     & & \\
     & & \Gamma \left( \dfrac{3 p + 2}{12} \right) \Gamma \left( \dfrac{3 p + 22}{12} \right) \nu^{-(p + 4) / 2}, \label{eqnKsyn}
\end{array}
\end{equation}
where $\Gamma(x)$ represents the $\Gamma$ function. Introducing the optical depth $\tau_\nu = k_\nu^{syn} R$, where $R$ is a typical size for the source, the emerging spectrum will be:
\begin{equation}
    \left\{
    \begin{array}{ll}
        I_\nu^{syn} = R j_\nu^{syn} & \mathrm{for} \quad \tau_\nu << 1 \\
         & \\
        I_\nu^{syn} = \dfrac{j_\nu^{syn}}{k_\nu^{syn}} (1 - \mathrm{e}^{-\tau_\nu}) & \mathrm{otherwise.}
    \end{array} \right. \label{eqnSynSpec}
\end{equation}
Integrating Eq.~\ref{eqnJsyn} over the particle energy distribution, leads to the well known result of a spectrum $I_\nu^{syn} \propto \nu^{5/2}$, in the optically thick domain, and to $I_\nu^{syn} \propto \nu^{-(p - 1) / 2}$ in the optically thin region.

On the other hand, a distribution of relativistic particles that interacts with a radiation field will modify the frequency spectrum of the incoming photons via Compton scattering. The expected emission coefficient can be expressed as:
\begin{equation}
    j_\nu^{IC} = h \nu c \int_{\nu_0}^{\nu_1} \de \nu' n(\nu') \int_{\gamma_{min}}^{\gamma_{max}} \de \gamma \dfrac{\de \sigma_{KN}}{\de \nu'} N(\gamma), \label{eqnJic}
\end{equation}
where $n(\nu')$ represents the density of incoming photons, $\nu_0$ and $\nu_1$ are the boundaries of the seed spectrum frequency range and $\de \sigma_{KN} / \de \nu'$ is the differential Klein-Nishina cross-section for scattering between an initial frequency $\nu'$ and a final frequency $\nu$. In our case, the seed spectrum is the synchrotron radiation field, computed in Eq.~(\ref{eqnSynSpec}). The corresponding photon density can be calculated as:
\begin{equation}
    n(\nu') = \dfrac{4 \pi I_{\nu'}}{h \nu' c}. \label{eqnPhDens}
\end{equation}
The scattering cross-section for isotropic incident angles was evaluated by \citet{Jones68} and expressed in its differential form by \citet{Band85} and by \citet{Massaro06}:
\begin{equation}
    \begin{array}{rcl}
        \dfrac{\de \sigma_{KN}}{\de \nu'} & = & \dfrac{3 \sigma_{Th}}{16 \gamma^2 \nu'} \left[ 2 \eta \ln \eta + (1 + 2 \eta) (1 - \eta) + \dfrac{}{} \right. \\
        & & \\
        & & \left. \dfrac{1}{2} (1 - \eta) \dfrac{(4 h \nu' \eta / m c^2)^2}{1 + 4 h \nu' \eta / m c^2} \right],
    \end{array}
\end{equation}
where $\sigma_{Th}$ is the Thomson cross-section, while the factor $\eta$\ is defined as:
\begin{equation}
    \eta = \dfrac{\nu}{4 \gamma^2 \nu' (1 - h \nu / \gamma m c^2)}
\end{equation}
and the calculations are carried out in the limits:
\begin{equation}
    \left\{
    \begin{array}{l}
        1 \leq \dfrac{\nu}{\nu'} \leq \dfrac{4 \gamma^2}{1 + 4 h \nu' / m c^2} \\
        \\
        0 < \eta < 1.
    \end{array} \right. \nonumber
\end{equation}
Thus, in the simplifying assumption of a single Compton scattering, we finally obtain an estimate of the expected Compton spectrum, in the form of:
\begin{equation}
    I_\nu^{IC} = R j_\nu^{IC}. \label{eqnICspec}
\end{equation}

\begin{figure}
    \centering
    \includegraphics[width=0.45\textwidth]{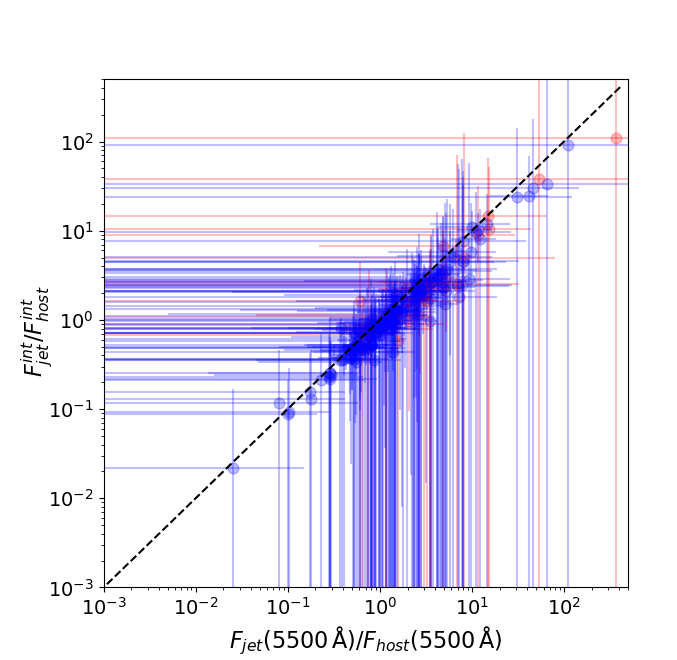}
    \caption{Comparison between the ratio among jet and host galaxy flux calculated at $5500\,$\AA\ and as a result of flux integration over the $3500\,$\AA\ -- $8000\,$\AA\ rest frame wavelength range. Blue points represent objects with an acceptable spectral fit, while red points are objects with a rejected solution. The black dashed line is the identity relation. \label{figRatioComparison}}
\end{figure}
\begin{figure*}
    \centering
    \includegraphics[width=0.9\textwidth]{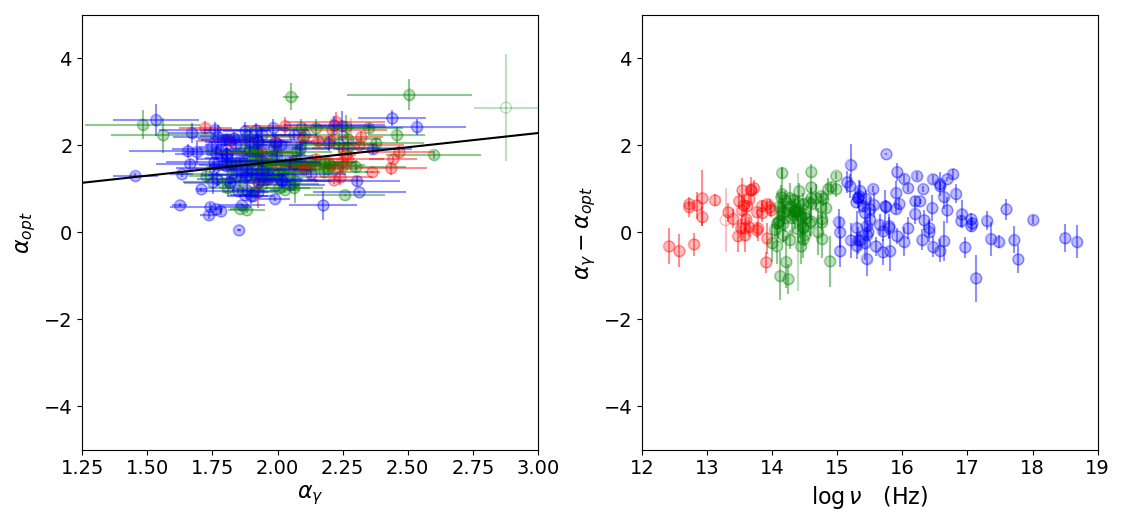}
    \caption{Same as in Fig.~\ref{figSpecInd}, but without accounting for the host galaxy component in the measurement of the optical spectral index. \label{figSpecIndNoHost}}
\end{figure*}
\section{Flux ratio comparison}
\label{secCompareRatios}
As a consistency check on the flux ratio measurements, we compared the results of the estimate carried out at a rest frame wavelength of $5500\,$\AA\ \citep[following the approach of][]{Goldoni21} and by integrating the jet and host galaxy flux in the $3500\,$\AA\ -- $8000\,$\AA\ wavelength range. The result is illustrated in Fig.~\ref{figRatioComparison}. We observe that the two approaches give very similar results, as long as the jet power is less than 30 times stronger than the host component. In any case, given the large uncertainties of the ratio, particularly when the host component is weak, all measurements are consistent with an identity relation.

\section{Single power-law fits}
\label{secFitNoHost}
The relation between the optical and $\gamma$-ray spectral indices, without accounting for the host galaxy component, is plotted in Fig.~\ref{figSpecIndNoHost}, with the same symbology as in Fig~\ref{figSpecInd}. Although we can still observe some degree of similarity, due to approximately half of the sample being dominated by the jet, rather than the host, there is no clear difference in the optical index of HSP, ISP and LSP objects and the optical and $\gamma$-ray spectral indices do not appear correlated within their uncertainty range. The optical spectral indices are generally softer than in the host subtracted case and the spectral index difference is reduced to an average value of $\langle \alpha_\gamma - \alpha_{opt} \rangle = 0.483$. In this case, the best fit relation is described by:
\begin{equation}
    \alpha_{opt} = (0.654 +/- 0.110) \alpha_\gamma + (0.324 +/- 0.445),
\end{equation}
with a weighted correlation coefficient $R = 0.518$ and a null hypothesis probability $p_0 = 0.99999$.
\end{document}